\documentclass[acmsmall]{acmart}

\AtBeginDocument{%
  \providecommand\BibTeX{{%
    \normalfont B\kern-0.5em{\scshape i\kern-0.25em b}\kern-0.8em\TeX}}}

\acmJournal{PACMHCI}
\acmYear{2022} \acmVolume{6} \acmNumber{CSCW1} \acmArticle{87} \acmMonth{4} \acmDOI{10.1145/3512934}

\usepackage{numprint}

\usepackage[inline]{enumitem}
\usepackage{listings}
\usepackage{booktabs}
\usepackage{tabularx}
\usepackage{multirow}

\usepackage{caption}

\usepackage{xcolor, colortbl}
\definecolor{table_blue}{RGB}{53, 132, 187}
\definecolor{table_orange}{RGB}{255, 139, 38}
\definecolor{table_green}{RGB}{65, 169, 65}

\usepackage{tikz}
 
\usepackage{enumitem}

\newcommand{\fullcite}[1]{\citeauthor{#1}~\cite{#1}}

\newif\ifdiff

\newcommand{\cBlue}[1]{\ifdiff{\leavevmode\color{blue}{#1}}\else{#1}\fi}
\newcommand{\cBlueSimple}{\ifdiff\color{blue}\else{}\fi}
\newcommand{\cBlueCaption}{\ifdiff\captionsetup{labelfont={color=blue},font={color=blue}}\else{}\fi}

\newcommand{\cTeal}[1]{\ifdiff{\leavevmode\color{teal}{#1}}\else{#1}\fi}

\difffalse

\newcommand{\filan}{P1} 
\newcommand{\fuma}{P2} 
\newcommand{\rocco}{P3} 
\newcommand{\plamen}{P4}
\newcommand{\sosi}{P5} 
\newcommand{\legna}{P6}
\newcommand{\bakke}{P7}
\newcommand{\andre}{P8}
\newcommand{\ciuff}{P9} 
\newcommand{\dani}{P10} 
\newcommand{\piemo}{P11} 
\newcommand{\giulia}{P12} 
\newcommand{\matt}{P13} 
\newcommand{\doss}{P14} 
\newcommand{\dista}{P15} 
\newcommand{\sann}{P16} 
\newcommand{\gabri}{P17} 
\newcommand{\sagli}{P18} 
\newcommand{\cordi}{P19} 
\newcommand{\lanott}{P20} 
\newcommand{\tess}{P21} 
\newcommand{\consort}{P22} 

\begin{document}

\title{Eliciting Best Practices for Collaboration with Computational Notebooks}

\author{Luigi Quaranta}
\orcid{0000-0002-9221-0739}
\affiliation{%
 \institution{University of Bari}
 \city{Bari}
 \country{Italy}
}
\email{luigi.quaranta@uniba.it}

\author{Fabio Calefato}
\orcid{0000-0003-2654-1588}
\affiliation{%
 \institution{University of Bari}
 \city{Bari}
 \country{Italy}
}
\email{fabio.calefato@uniba.it}

\author{Filippo Lanubile}
\orcid{0000-0003-3373-7589}
\affiliation{%
 \institution{University of Bari}
 \city{Bari}
 \country{Italy}
}
\email{filippo.lanubile@uniba.it}



\begin{abstract}
Despite the widespread adoption of computational notebooks, little is known about best practices for their usage in collaborative contexts.
In this paper, we fill this gap by eliciting a catalog of best practices for collaborative data science with computational notebooks.
With this aim, we first look for best practices through a multivocal literature review. Then, we conduct interviews with professional data scientists to assess their awareness of these best practices.  Finally, we assess the adoption of best practices through the analysis of \numprint{1,380} Jupyter notebooks retrieved from the Kaggle platform.
Findings reveal that experts are mostly aware of the best practices and tend to adopt them in their daily work. Nonetheless, they do not consistently follow all the recommendations as, depending on specific contexts, some are deemed unfeasible or counterproductive due to the lack of proper tool support.
As such, we envision the design of notebook solutions that allow data scientists not to have to prioritize exploration and rapid prototyping over writing code of quality.
\end{abstract}



\maketitle

\section{Introduction}

\cBlue{Data science is a multi-disciplinary, collaborative process where  different stakeholders with varied backgrounds and skillsets are involved in a pipeline of complex activities ranging from data preparation and modeling to deployment and monitoring~\cite{muller_how_2019}.
To support these practices, data scientists rely on a variety of tools such as file-sharing utilities, project management software, (a)synchronous discussion apps, and presentation suites \cite{zhang_how_2020}.
However, computational notebooks, and in particular their most popular implementation, Jupyter Notebook \cite{perkel2018jupyter}, have become the tool of choice for data scientists  \cite{rule_exploration_2018}.}
The compelling interactive features of computational notebooks have shown to be particularly convenient for data-centric programming tasks, such as data exploration and statistical data analysis \cite{kluyver2016jupyter, toomey2017jupyter}, as well as for model prototyping \cite{wang_how_2019, toomey2017jupyter}. Moreover, the self-documenting format that computational notebooks provide has made them successful instruments for the communication and dissemination of analytical results \cite{perez_project_2015, rule_aiding_2018, rule_exploration_2018, rule2019ten}.

The growing popularity of computational notebooks has attracted the interest of many researchers (e.g., \cite{wang_better_2020,wang_restoring_2020, rule_exploration_2018}). Besides the clear advantages of providing a narrative through code cells interleaved with inline documentation, many pitfalls and downsides have been identified when using notebooks \cite{chattopadhyay2020s, grus_i_2018}, particularly in collaborative environments or when they are developed as production-level artifacts within AI-based systems \cite{pimentel_large-scale_2019, rule_aiding_2018, kery2018interactions, koenzen_code_2020, head_managing_2019}.

There are two typical scenarios in the usage of computational notebooks \cite{rule_exploration_2018}. 
\cBlue{The first scenario concerns \textbf{\textit{individual}} sessions of work where data scientists create `disposable' notebooks to rapidly explore datasets and use the resulting insights to inform further activities for themselves or their team; 
because this scenario does not entail collaboration, ensuring that notebooks contain high-quality code and documentation is not a priority.} 
On the other hand, the second scenario involves sharing analytical insights with others. 
\cBlue{\fullcite{wang_human-ai_2019} describe this \textbf{\textit{collaborative}} scenario as a `scatter-gather' process where data scientists switch repeatedly between individual exploratory analyses (scatter) and insights discussion sessions (gather) until project completion.} 
As such, here data scientists should make sure to fully leverage the self-documenting nature of computational notebooks, by writing exhaustive explanations of their work in natural language text \cite{rule_exploration_2018} and checking that their computation can be re-executed without errors and fully reproduced by other colleagues \cite{pimentel_large-scale_2019, wang_restoring_2020}.

\cBlue{In this paper, we study computational notebooks developed in the aforementioned collaborative scenario to investigate the availability and adoption of best practices for writing, organizing, and sharing notebooks in \cTeal{the context of professional} data science teams.
To this aim, we start }by asking the following research question:

\begin{description}

    \item[RQ1] \textit{\cTeal{What are the} best practices for collaboration with computational notebooks?}
    
\end{description}

To answer RQ1, we perform a multivocal literature review \cite{garousi_guidelines_2019}, a form of systematic literature review that includes as sources not only peer-reviewed papers published in journals or conference proceedings (white literature) but also informal documents (grey literature), like whitepapers or blog posts, produced by professionals based on their practical experience. 

The output of this literature review is a catalog of 17 best practices for collaboration with computational notebooks. 
As neither white literature nor grey literature provides any empirical evidence of how popular best practices are,  
we also ask:

\begin{description}

    \item[RQ2] \textit{What collaboration-specific best practices do experienced data scientists actually follow  \cBlue{when working with computational notebooks}}?
    
\end{description}

To answer RQ2, we first perform a qualitative analysis based on interviews with 22 practitioners from different companies.
Then, we conduct an archival study to quantitatively assess the overall adoption of notebook best practices that involve collaboration. 
Accordingly, we build and analyze a dataset of \numprint{1,380} notebooks, available on the Kaggle platform, which have been collaboratively created by experienced data scientists.

We found that practitioners are generally aware of most of the best practices in the literature, although they do not always apply the recommended behaviors, deeming them unfeasible or counterproductive in definite contexts due to the lack of proper tool support. Therefore, we identify some possible improvements to the notebook development environments, such as extensions to support code modularization, linting, and testing.

The original contribution of this study is twofold.
First, we provide a validated catalog of best practices for collaboration with computational notebooks. This contribution might inform practitioners willing to improve how they use computational notebooks.
Second, we highlight shortcomings in notebook development environments that might hinder the adoption of specific best practices.

The remainder of this paper is organized as follows.
In Section~\ref{sec:related_work}, we \cBlue{review the literature on collaboration in data science and tools supporting data science work, with a focus on computational notebooks  and their most popular implementation, Jupyter Notebook}.
In Section~\ref{sec:MLR}, we present the multivocal literature review.
In Section~\ref{sec:qualitative_study}, we report the qualitative assessment of emerged best practices, based on interviews with experts.
Next, in Section~\ref{sec:quantitative_study}, we delve into the quantitative study on the adoption of best practices. 
All the results of this study are discussed in Section~\ref{sec:discussion}, along with the identified limitations, and compared to relevant prior work.
Finally, we conclude the paper in Section~\ref{sec:conclusion}.

\section{Related Work}
\label{sec:related_work}

\cBlue{In this section, to contextualize our study, we draw upon prior work focused on the practices and tools used in collaborative data science.}

\subsection{\cBlue{Collaboration in Data Science}}
\label{sec:collaboration_in_data_science}

\cBlue{Data science is a multidisciplinary `team sport'~\cite{wang_how_2019,kim_emerging_2016} that involves many different activities and stakeholders, such as scientists and engineers who clean data and train models with the help of domain experts, and operations team members who take care of model deployment and monitoring.
\fullcite{muller_how_2019} proposed the term `data science worker' to capture the broad diversity of the job categories and titles involved. 
Due to this multidisciplinary nature, collaboration in data science can be challenging.

In contrast with software engineering research, which has built strong literature on collaboration practices in software development (e.g., \citep{herbsleb_global_2007,bird_sociotechnical_2011,calefato_agile_2019}), only recently researchers have started focusing on the practices and challenges within data science teams (e.g., \cite{zhang_how_2020, passi_trust_2018, aragon_developing_2016, chancellor_relationships_2019}). 
Translating data science results into business actions requires successful communication between data scientists and other, non-technical  stakeholders who typically lack a common ground and a shared vocabulary~\cite{passi_trust_2018}. 
\fullcite{matsudaira_science_2015} observed that different teams within a company --- or even team members with different roles --- seldom speak the same language.
A similar experience is reported by 
\fullcite{kandel_enterprise_2012} and \fullcite{hou_hacking_2017}, who found that collaborations are rare or difficult for data science teams working on marketing, finance, and civic data hackathons due to the diversity of tools and languages. 
In other words, successful communication of data science results goes beyond simply presenting model performance metrics and requires, instead, taking into account also customer experience and business metrics.

Furthermore, challenges in collaborative data science can be different from those faced in conventional software engineering. For example, big-data sharing and management create a wealth of new issues that are specific to data science teams~\cite{bopp_disempowered_2017}.

Due to the steadily growing demand for data analysis~\cite{davenport2012data}, ensuring efficient collaboration within data science teams is becoming increasingly important as well as challenging. 
As such, with this study, we aim to identify the existing best practices of data science collaboration reported in the literature, focusing specifically on computational notebooks, and the presence of mismatches and gaps as compared to the best practices actually followed by practitioners in the wild.
}




\subsection{\cBlue{Data Science Tools: Computational Notebooks}}
\label{sec:computational_notebook_platforms}

\cBlue{
Tools have a direct impact on collaborative practices.
\fullcite{zhang_how_2020} surveyed 183 data scientists and found that data science teams are extremely collaborative at all the stages of the adopted  workflow, although the extent varies according to the stakeholders involved and the tools they use. 
As such, it is important to track the usage and evolution of tool support, especially in the data science field  where new tools --- powered by state-of-the-art AI technologies --- emerge every day.

Many systems have been proposed to support the various aspects of data science work practices.
Data exploration is one of the most common and often time-consuming activities in data science work.
Nowadays, data scientists can count on many visual analytics tools designed to help them collaboratively explore and make sense of large datasets.
Popular examples are TensorBoard\footnote{\url{https://www.tensorflow.org}} and Tableau.\footnote{\url{https://www.tableau.com}}

Another category of tools that is rapidly gaining ground is focused on the tracking and reproducibility of data science experiments.
Reproducing experiments that involve machine learning techniques can be challenging, due to the many steps that compose the typical AI/ML workflow \cite{amershi_software_2019}.
To ensure reproducibility, data scientists can leverage solutions like MLflow,\footnote{\url{https://mlflow.org}} a popular open-source experiment tracker and model registry, and DVC,\footnote{\url{https://dvc.org}} a data versioning tool capable of recording processing steps and representing them as a reproducible pipeline.

Researchers and practitioners have long attempted to create a common digital format that would allow data scientists to present, reproduce, and share --- in a word, collaborate at --- data analysis results.}
Back in 1984, Donald Knuth conceived \textit{literate programming} \cite{knuth_literate_1984}, an innovative paradigm for software development in which code is written side by side with the explanation of its logic; programs would appear as peculiar text documents, with natural language excerpts -- delivering the code's rationale -- interleaved by macros and traditional code fragments.
Clear explanations in plain natural language were also supposed to enhance collaboration, improving the overall comprehensibility of programming artifacts. 
Some decades later, the literate programming paradigm  has become the conceptual base for modern \textit{notebook interfaces}, commonly known as \textit{computational notebooks}. 

Computational notebooks are interactive documents organized in cells of three different types: \textit{markdown}, \textit{code}, and \textit{raw}.
Raw cells are used to store instructions for external conversion tools. Markdown cells contain natural language excerpts that define the notebook narrative.
Code cells, which can be written in several programming languages like Python, R, and Julia, are immediately followed by their output (e.g., text, images, tables), returned by an underlying kernel (interpreter).
Once executed, a code cell and its output are assigned the current value of the \textit{execution counter}, an internal variable that is incremented every time the kernel is successfully invoked on the notebook. 
Furthermore, users can choose to store executed or non-executed notebooks: in the latter case, only \textit{prospective data} are saved (i.e., markdown cells and code cells without their output), in the former, \textit{retrospective data} (i.e., the output of code cells and their execution counts) are also stored \cite{freire2008provenance}.

\cBlue{There are many examples of computational notebook platforms.
Some of them are language-specific, e.g., RStudio\footnote{\url{https://www.rstudio.com}} (using R Markdown), Wolfram Notebooks\footnote{\url{https://www.wolfram.com/notebooks/}} (using Wolfram's proprietary languages), and Observable\footnote{\url{https://observablehq.com}} (using Javascript).
Other platforms allow multiple programming languages, e.g., Apache Zeppelin\footnote{\url{http://zeppelin.apache.org}} and Spark Notebook;\footnote{\url{https://github.com/spark-notebook/spark-notebook}} one of them, Polynote,\footnote{\url{https://polynote.org}} even supports mixing different languages within the same notebook.}
To date, the most widely adopted computational notebook systems are \textit{Jupyter Notebook} and \textit{Jupyter Lab}.\footnote{\url{https://jupyter.org}}
Jupyter Notebook is an open-source web application for the creation of notebooks that comprise live code, equations, visualizations, and narrative text.
Jupyter Lab, its evolution, is a full-fledged web-based IDE that can handle notebooks as well as code, data, and terminal sessions, all in a single multi-paneled window. 
Both are maintained by the nonprofit organization Project Jupyter \cite{perez_project_2015}, allow the execution of code in multiple languages (e.g., Python, R, and Julia), share the notebook file format (\texttt{.ipynb}), and can be collaboratively accessed using their multi-user version, \textit{JupyterHub}.

\cBlue{Recently, there have been several efforts aimed at improving Jupyter Notebook for collaboration within data science teams and beyond the perspective of the individual data scientist.}
A noteworthy example is \textit{Google Colaboratory}\footnote{\url{https://colab.research.google.com}} (also known as \textit{Colab}), a cloud service offering Google's implementation of the notebook interface backed by free-to-use computational power.
Since it shares part of Google Drive's backend, Colab offers its users state-of-the-art collaboration capabilities. 

Another popular Google service, leveraging computational notebooks, is Kaggle.\footnote{\url{www.kaggle.com}} 
The platform offers a versioned, cloud computational environment that allows the creation of scripts and computational notebooks in R and Python.
The main interface of Kaggle is a collaborative implementation of Jupyter Notebook, called \textit{Notebooks IDE},\footnote{\url{www.kaggle.com/docs/notebooks\#the-notebooks-environment}} which allows teammates to collaborate. 
Kaggle  hosts data science competitions for practitioners at all levels of expertise.
\cBlue{Deepnote\footnote{\url{https://deepnote.com}}, Jovian,\footnote{\url{https://www.jovian.ai}} Observable,\footnote{\url{https://observablehq.com}} and Databricks Collaborative Notebooks\footnote{\url{https://databricks.com/product/collaborative-notebooks}} are other notable examples of computational notebook platforms running in the cloud and offering real-time collaboration. 
Despite the obvious benefits of improved collaboration like shared context, encouraged exploration, and reduced communication costs, prior research~\cite{wang_how_2019} has found that synchronous notebooks can also lead to imbalanced participation and increased coordination costs without a proper collaboration strategy.

Jupyter notebooks have become the tool of choice for many data scientists working in a variety of settings, including research, industry, and education~\cite{perkel2018jupyter}.
They have become so popular that even traditional IDEs and advanced text editors (e.g., PyCharm,\footnote{\url{www.jetbrains.com/pycharm}} Microsoft Visual Studio Code),\footnote{\url{https://code.visualstudio.com}} now offer support, accounting for their growing spread in software repositories. 
Given this massive adoption, in our study we focus specifically on investigating the best practices for collaboration involving the development of Jupyter notebooks.
}

\subsection{\cBlue{Jupyter Notebook Criticism}}

\cBlue{
The popularity of Jupyter notebooks in the data science community has drawn the attention of many researchers who have identified a considerable number of limitations \cite{grus_i_2018}.

\fullcite{rule_exploration_2018} conducted a large-scale study by analyzing over 1 million open-source notebooks from GitHub and found that data scientists rarely take advantage of notebook literate programming capabilities, i.e., the ability to complement analytical code with explanatory text.
\fullcite{kery2018story} studied coding behaviors in the computational notebooks of 21 data scientists and reported their struggles with keeping track of experiment history.
Both studies report on an existing the tension between exploration and explanation in computational notebooks and suggest that data scientists seem to privilege the former (i.e., the individual `scatter' activities~\citep{wang_human-ai_2019}) in their day-to-day work, thus writing informal notebooks that may be difficult to understand later on~\citep{rule_exploration_2018}.

Several other studies have proposed the use of Jupyter Notebook extensions to address `messy' notebooks~\cite{head_managing_2019}. 
For example, \fullcite{kery2018interactions} added a local versioning feature to help data scientists keep track of experimental runs.
To encourage easier exploration during complex analytical tasks, \fullcite{rule_aiding_2018} developed an extension that enables lightweight cell folding and annotation.
\fullcite{wang_callisto_2020} proposed a chat service integrated into Jupyter Notebook, which automatically captures and stores contextual links between discussion messages and notebook elements, thus helping to make sense of the overall process that led to the creation of notebooks.
Inspired by \textit{program slicing} \cite{weiser_program_1984}, \fullcite{head_managing_2019} developed an interactive tool suite that allows the extraction of specific lines of code that produce a certain output from a messy notebook to a new notebook or the clipboard.

\cTeal{Not} only documentation (or lack thereof) but also the quality of code contained in Jupyter notebooks has raised concerns.
\fullcite{pimentel_large-scale_2019} performed a large-scale study about the quality and reproducibility of Jupyter notebooks.
They collected 1.4 million notebooks from GitHub and performed a comprehensive analysis by assessing common practices in notebook usage (e.g., the use of literate programming features) as well as the presence of typical `smells' in notebook execution (e.g., out-of-order cells, execution counter skips as a clue of hidden states, etc.);
then, they  attempted the re-execution of a subset of the selected Python notebooks and succeed without errors in only $24\%$ of the cases.
\fullcite{wang_better_2020} evaluated the quality of code contained in notebooks gathered from a gallery maintained by Project Jupyter.
Their study shows that --- despite being select artifacts, many of which with an educational purpose --- these notebooks contain a large amount of poor-quality code. 
Conformance to the Python coding best practices was assessed using a tool that checked for the presence of violations of the coding conventions defined in PEP~8,\footnote{\url{www.python.org/dev/peps/pep-0008}} the official `Style Guide for Python Code' used to develop the standard library in Python distributions, as well as the presence of unused variables and deprecated functions.
\fullcite{koenzen_code_2020} analyzed code duplication and reuse in Jupyter notebooks.
Albeit duplicating  code is generally considered a `bad smell' causing software maintainability issues \cite{fowler_refactoring_2018,tufano_when_2015,juergens_code_2009}, it is an expedient often used in data science for writing code faster and iterating over data to expedite experimentation  \cite{beth_kery_exploring_2017, brandt_opportunistic_2008}. 
\citeauthor{koenzen_code_2020}  found that, on average, about 8\% of code in GitHub repositories containing Jupyter notebooks is self-duplicated, arguing that notebook authors might be more concerned with ease of use than coding best practices.

Finally, \fullcite{rule2019ten} shared a set of ten rules for writing and sharing computational analysis in Jupyter notebooks.
These rules cover the whole notebook lifecycle and address a wide range of issues, from writing good documentation in narrative form, with a specific audience in mind, to modularizing code and using version control.

All these studies suggest that --- to successfully support data science work --- the use of notebooks should be disciplined and informed with shared best practices.
Albeit not validated, the papers by \fullcite{rule2019ten} and \fullcite{pimentel_large-scale_2019} propose a list of recommendations for notebook writing.
In this work, we take a step forward by (i) systematically collecting best practices for the collaborative use of computational notebooks (including those put forward by these two studies) and (ii) assessing how practitioners perceive recommendations and adopt them in their daily work.

}

\section{Multivocal Literature Review}
\label{sec:MLR}

Multivocal literature review (MLR) is a form of systematic review that includes documents from the so-called `grey' literature (e.g., blog posts and whitepapers) in addition to formal peer-reviewed papers (`white' literature).

Garousi et al.~\cite{garousi_need_2016,garousi_guidelines_2019,garousi_benefitting_2020} showed the benefits of complementing \cBlue{Systematic Literature Reviews (SLRs)} with grey literature in software engineering research. 
Software engineering practitioners typically lack the time or expertise to access academic publications; thus, they mainly rely on grey literature to stay up to date and inform their work. For the same reasons, they also often share their ideas and experiences in grey literature form. 
The resulting wealth of information remains off the radar of researchers when they limit themselves to considering white literature in their studies.

\cBlue{
Assuming that a similar situation must hold in the data science field, we decided to include grey literature in our review.
In this section, we briefly summarize our steps to perform the MLR. For a more detailed description of the process, please refer to Appendix~A.
}

\subsection{Method}

In March 2020, we performed a \cBlue{systematic} search on five different search engines. 
To retrieve white literature (i.e., scientific papers), we searched four digital libraries: in particular two broad-coverage abstract databases, i.e., Google Scholar and Web of Science, and two limited-coverage full-text databases, i.e., IEEE Xplore and ACM Digital Library. To retrieve grey literature, we performed a generic web search using Google. \cBlue{In the following, we describe how we identified a suitable set of search terms, selected relevant results, complemented our search with the help of the snowballing technique, and finally analyzed the resulting corpus to derive a catalog of best practices.

\subsubsection{Identification of the search terms} As the first step of our MLR, we identified a set of suitable search terms.
We started by considering an initial set of words that we had already encountered in the literature on computational notebooks, i.e., `computational notebook,' `Jupyter Notebook,' `best practice,'' and `recommendation.'
Next, we expanded the set by including relevant synonyms of the term `recommendation', i.e., `advice,' `tip,' `hint,' `guidance,' `guideline,' and `hack.'
With the resulting set of terms, we built a tentative query and performed a search on IEEE Xplore and ACM Digital Library, thus retrieving an initial set of white literature results. From these, we extracted the author-defined keywords, seeking possible neglected search terms. However, we were not able to find any new relevant keyword and thus accepted the tentative set of search terms as final.

We then composed the following query, using a generic syntax, from which we created specialized instances to account for the syntactic requirements of each specific search engine:\footnote{In the search-engine-specialized versions of the query, we explicitly included countable search terms in both their singular and plural forms.}
}




{\cBlueSimple
\begin{lstlisting}

("best practice" OR "guideline" OR "recommendation" OR "advice" OR "tip" 
    OR "hint" OR "guidance" OR "hack") 
AND 
("Jupyter Notebook" OR "Jupyter notebooks" OR "computational notebook")

\end{lstlisting}
}

\cBlue{
\subsubsection{Results sampling} Table~\ref{tab:number-of-results} summarizes the number of search results retrieved from each search engine and specifies where we searched for the keywords (i.e., title, abstract, metadata, etc.).
Academic digital libraries returned a small number of search hits (IEEE Xplore, ACM Digital Library, and Web of Science returned 3, 11, and 17 results respectively), which we completely and thoroughly reviewed, with the only exception of Google Scholar, which returned 7,610 results.
Not surprisingly, the generic Google search returned a far greater number of results (2,740,000).

To manage the extensive number of hits retrieved from Google Scholar and Google Search, we sampled the results following two different approaches.
In the case of Google Scholar, we went through the first 366 search hits (i.e., 37 10-item pages) --- the same size as a representative sample of the complete result set with a 5\% margin of error and 95\% confidence level.
In the case of Google Search, we checked the relevance of the first 100 hits (i.e., ten 10-item pages), and stopped at the 10th page after inspecting six consecutive pages of unrelated results.

\begin{table}[tb]
\cBlueSimple
\cBlueCaption
\small
\begin{tabular}{@{}lll@{}}
\toprule
\textbf{Search Engine}       & \textbf{Items searched}                                        & \textbf{No. of search results} \\ \midrule
IEEE Xplore         & All metadata                                             & 3                        \\
ACM Digital Library & Publication title, abstract, and author-defined keywords & 11                       \\
Google Scholar      & Google Scholar default search                            & 7,610                     \\
Web of Science      & Publication title, abstract, and author-defined keywords & 17                       \\
Google Search       & N.A.            & 2,740,000 \\
\bottomrule
\end{tabular}

\caption{Number of search results obtained from each search engine.}
\label{tab:number-of-results}
\vspace{-5mm}
\end{table}

\subsubsection{Selection criteria}

In our MLR, we decided to include only written articles; accordingly, we discarded search results related to other multimedia types, such as videos, podcasts, etc. Moreover, we kept only articles written in English whose publication date was after the birth of Project Jupyter (2015), which popularized computational notebooks.

As for the content, we kept only articles containing a list of best practices for the collaborative use of computational notebooks in a generic professional context. Consequently, we did not gather articles providing introductory guides on computational notebooks, low-level productivity tips for notebook writing, or recommendations on using specific tools or notebook extensions. Furthermore, we did not consider articles addressed to a specific domain (e.g., best practices for the use of computational notebooks in education).
For further details on the rationale for the selection criteria, please refer to Table~\ref{tab:search-results-criteria} in Appendix~A.

\subsubsection{Snowballing} As the application of our filtering criteria led to an overall low number of selected papers, we employed the snowballing technique to improve the coverage of academic articles.
In particular, we engaged in a backward and forward snowballing process \cite{wohlin_guidelines_2014}, by  leveraging respectively the references in and the citations of the selected white literature.

Our start set included the only two academic papers that we had been able to retrieve \cite{rule_ten_2018, pimentel_large-scale_2019}. We identified one more article \cite{woodbridge_jupyter_2017} via backward snowballing, while the forward snowballing produced no further results. Incidentally, the only new reference was a grey literature article (i.e., a blog post) that did not lead to further items on which to iterate the snowballing process.}

\subsubsection{\cBlue{Data extraction}}

\cBlue{After complementing the data collection step with snowballing, we were left with a set of 14 articles to analyze (3 white literature papers and 11 grey literature articles).
W}e carefully read each \cBlue{of them, extracting excerpts that described one or more recommendations; in doing so, we built a preliminary} raw catalog of best practices.

\cBlue{
By `best practice' for collaboration with computational notebooks, we specifically intend an optimal, high-level, and general behavior that can be adopted to improve how data scientists work together around computational notebooks. Based on this definition, we defined a list of inclusion/exclusion criteria and filtered the raw catalog accordingly.
In particular, and consistently with the paper selection criteria, we excluded from the final catalog all the best practices recommending the use of external tools (e.g., the PyCharm IDE) or specific Jupyter Notebook extensions (e.g., toc2);\footnote{\url{https://jupyter-contrib-nbextensions.readthedocs.io/en/latest/nbextensions/toc2/README.html}} the use of specific IPython magic functions or specific Python packages; low-level productivity tips (e.g., ``learn the Jupyter keyboard shortcuts''); all the opinionated recommendations that were not supported by proper argumentation (e.g., ``use Jupyter Lab instead of Jupyter Notebook''); the best practices that provide generally valid coding advice (e.g., ``Use meaningful variable names'').

For the rationales behind the best-practice selection criteria, please refer to Table~\ref{tab:best-practices-criteria} in Appendix~A.} 

\subsubsection{\cBlue{Data analysis}}
To further refine our catalog, we performed a thematic analysis of the raw collection of best practices \cite{clarke2015thematic}. \cBlue{In particular, w}e assigned an open set of codes to each best practice \cBlue{and, upon consolidation of the code set, we performed a focused coding, noting down the support of each identified code} (i.e., the number of sources in which they appeared); finally, we grouped the codes into common themes.

\subsection{Results}

A summary of the selected search results is available in Table~\ref{tab:search-results}. 
As the third article [W3] retrieved via Google Scholar is merely a pre-print version of the first one [W1], the actual number of academic articles retrieved through all four academic search engines is \cBlue{three} [W1, W2\cBlue{, W4}].
On the other hand, the overall number of selected Google Search results is \cBlue{thirteen}; however, one article [W1] had been already retrieved via Google Scholar. 
Moreover, two \cBlue{pairs of} blog posts \cBlue{(i.e., }[G3, G4] \cBlue{and [G6, G12])} have the same content. Therefore, the actual count of non-academic articles retrieved via Google Search is \cBlue{ten}.
\cBlue{Finally, the snowballing process returned one more article, a technical blog post [G13], raising the number of grey literature items to eleven.}

\begin{table}[tb]
\footnotesize
{\def\arraystretch{1.5}
\begin{tabular}{@{}llp{10cm}@{}}
\toprule
\textbf{Search engine} & \textbf{ID} & \textbf{Results} \\ \midrule
 
\multirow{3}{*}{Google Scholar} 
 & [W1] & Rule, A. et al. ``Ten Simple Rules for Writing and Sharing Computational Analyses in Jupyter Notebooks. \cite{rule2019ten} \\
 & [W2] & Pimentel, JF et al. ``A Large-Scale Study about Quality and Reproducibility of Jupyter Notebooks.'' \cite{pimentel_large-scale_2019} \\
 & [W3] & Rule, A. et al. ``Ten Simple Rules for Reproducible Research in Jupyter Notebooks.'' \cite{rule_ten_2018} \\ 
 & \cBlue{[W4]} & \cBlue{Beg, Marijan et al. ``Using Jupyter for Reproducible Scientific Workflows'' \cite{beg_using_2021-1}} \\
 \midrule
 
IEEE Xplore
 & [W2] & Pimentel, JF et al. ``A Large-Scale Study about Quality and Reproducibility of Jupyter Notebooks.'' \cite{pimentel_large-scale_2019} \\ 
 & \cBlue{[W4]} & \cBlue{Beg, Marijan et al. ``Using Jupyter for Reproducible Scientific Workflows'' \cite{beg_using_2021-1}} \\
 \midrule

ACM Digital Library
 & [W2] & Pimentel, JF et al. ``A Large-Scale Study about Quality and Reproducibility of Jupyter Notebooks.'' \cite{pimentel_large-scale_2019} \\ \midrule

Web of Science
 & & / \\ \midrule

\multirow{9}{*}{Google} 
 & [G1] & Haitz, Dominik. ``Jupyter Notebook Best Practices'' - Towards Data Science. \cite{dominik_haitz_jupyter_2019} \\
 & [G2] & ``Jupyter Notebook Manifesto: Best practices that can improve the life of any developer using Jupyter notebooks'' - Google Cloud Blog. \cite{michael_cheng_jupyter_2019} \\
 & [G3] & Karakan, Burak. ``Jupyter Notebook Best Practices'' - Level Up Coding. \cite{burak_karakan_jupyter_2020} \\
 & [G4] & ``Best Practices for Jupyter Notebooks'' - Saturn Cloud. \cite{saturn_cloud_dev_team_best_2020} \\
 & [G5] & ``Jupyter Notebook: The Definitive Guide.'' - DataCamp Community. \cite{karlijn_willems_jupyter_2019} \\
 & [G6] & ``Jupyter Notebook Best Practices for Data Science'' - KDnuggets. \cite{jonathan_whitmore_jupyter_2016} \\
 & [G7] & Kierzkowski, Roman. ``10 tips on using Jupyter Notebook'' - Medium. \cite{roman_kierzkowski_10_2017} \\
 & [G8] & ``Working efficiently with Jupyter Notebooks'' - Inovex Blog. \cite{florian_wilhelm_working_2018} \\
 & \cBlue{[G9]} & \cBlue{Perrier, Alexis. ``Best practices when sharing your data analysis - Jupyter Notebooks'' - ``Alexis Perrier - Data Science'' blog. \cite{perrier_best_2020}} \\
 & \cBlue{[G10]} & \cBlue{Mani, Atma. ``Coding Standards for Jupyter Notebook'' - Esri blog. \cite{mani_coding_2018}} \\
 & \cBlue{[G11]} & \cBlue{Miranda, Lester James V. ``How to use Jupyter Notebooks in 2020 (Part 2: Ecosystem growth)'' - Lj Miranda personal blog. \cite{miranda_how_2020-1}} \\
 & \cBlue{[G12]} & \cBlue{Whitmore, Jonathan. ``Jupyter Notebook Best Practices for Data Science'' - Silicon Valley Data Science. \cite{whitmore_jupyter_2015}.} \\
 & [W1] & RULE, Adam, et al. Ten Simple Rules for Reproducible Research in Jupyter Notebooks. \cite{rule2019ten} \\ \midrule
 
\cBlue{Snowballing} 
 & \cBlue{[G13]} & \cBlue{Woodbridge, Mark. ``Jupyter Notebooks and reproducible data science'' - Mark Woodbridge personal blog. \cite{woodbridge_jupyter_2017}} \\
 
 \bottomrule
\end{tabular}
\caption{Results matching the inclusion criteria for the multivocal literature review}
\vspace{-4mm}
\label{tab:search-results}
\vspace{-5mm}
}
\end{table}

As a result of our coding activity on the recommendations available in the identified articles, in Table~\ref{tab:MLR-summary} we report the best practices in our catalog, grouped by theme, which specifically address the collaborative use of computational notebooks. The remainder of this section (and study) will focus on this subset of collaboration-specific best practices. All the other recommendations, such as \textit{Leverage automation through continuous integration/deployment} or \textit{Use as few global variables as possible} --- which are not collaboration-specific as they apply to individual notebook writing as well or coding in general --- are therefore excluded from our analysis.

\begin{table}[t]
\centering
{\footnotesize
\begin{tabular}{@{}p{3.5cm}p{4cm}ccc@{}}

\toprule

\textbf{Theme} & \textbf{Best practice} & \textbf{Support [source IDs]} \\ \midrule

\multirow{5}{3.5cm}{\textbf{Make your analysis traceable and reproducible}} 
 & BP1 \textit{Use version control} & \cBlue{8} [W1, G1, G2, G3, G5, G6, G8\cBlue{, G11}] \\
 & BP2 \textit{Manage project dependencies} & \cBlue{7} [W1, W2, G2, G7, G8\cBlue{, G9, G13}]  \\
 & BP3 \textit{Use self-contained environments} & \cBlue{6} [W1, G1, G2, G7, G8\cBlue{, G13}]  \\
 & BP4 \textit{Put imports at the beginning }& \cBlue{5} [W2, G5, G7\cBlue{, G9, G10}]  \\
 & BP5 \textit{Ensure re-executability}\newline\hphantom{BP5} (\textit{re-run notebooks top to bottom}) & \cBlue{6} [W1, W2,\cBlue{ W4}, G2\cBlue{, G9, G11}] \\
 
 \midrule
 
\multirow{5}{3.5cm}{\textbf{Write high-quality code (i.e., code that can be easily shared and reused)}}
 & BP6 \textit{Modularize your code} & \cBlue{7} [W1, W2, G1, G3, G7, G8\cBlue{, G11}] \\
 & BP7 \textit{Test your code} & \cBlue{5} [W2, G2, G3, G7\cBlue{, G9}]  \\
 & BP8 \textit{Name your notebooks\newline\hphantom{BP8} consistently} & 3 [W2, G5, G6] \\
 & BP9 \textit{Stick to coding standards} & \cBlue{4} [G1, G3, G5\cBlue{, G10}] \\
 & BP10 \textit{Use relative paths} & 1 [W2] \\
 
 \midrule
 
\multirow{3}{3.5cm}{\textbf{Leverage the literate programming paradigm}}
 & BP11 \textit{Document your analysis} & \cBlue{8} [W1, G1, G3, G5, G8\cBlue{, G9, G10, G11}]  \\
 & BP12 \textit{Leverage Markdown headings}\newline\hphantom{BP12} \textit{to structure your notebook} & \cBlue{6} [W1, W2, G1, G3\cBlue{, G9, G10}] \\
 
 \midrule
 
\multirow{2}{3.5cm}{\textbf{Keep your notebook clean and concise}}
 & BP13 \textit{Keep your notebook clean }& 3 [W1, W2, G3] \\
 & BP14 \textit{Keep your notebook concise} & \cBlue{4} [W1, G5\cBlue{, G9, G10}] \\
 
 \midrule
 
\textbf{Distinguish production and development artifacts}
 & BP15 \textit{Distinguish production}\newline\hphantom{BP15} \textit{and development artifacts} & \cBlue{5} [G2, G5, G6\cBlue{, G9, G11}] \\

\midrule

\multirow{2}{3.5cm}{\textbf{Embrace open dissemination}}
 & BP16 \textit{Make your notebooks available} & \cBlue{2} [W1\cBlue{, G9}]  \\
 & BP17 \textit{Make your data available} & \cBlue{4} [W1\cBlue{, G9, G11, G13}]  \\
 
\bottomrule

\end{tabular}
}
\caption{Catalog of the collaboration-specific best practices for notebook usage, extracted through the multivocal literature review}
\label{tab:MLR-summary}
\vspace{-5mm}
\end{table}

\subsubsection{Themes emerged} In this paragraph, we review the themes that emerged from the multivocal literature review; we define each related best practice, enriching explanations with the most informative quotes from the literature.

\textbf{Make your analysis traceable and reproducible.} Among the recommendations that have the largest support in the analyzed articles, many are concerned with the traceability and reproducibility of the analyses performed within computational notebooks. \cBlue{Eight} sources, from white literature and grey literature as well, recommend \textit{putting notebooks under version control}. This recommendation is often accompanied by tips on how to mitigate the readability issue in Jupyter Notebook diffs, which ``\textit{are expressed as changes in the abstruse JSON metadata for the notebook}''  [W1].
\cBlue{One solution coming at the expense of notebook completeness is to remove the notebook output before committing changes:} ``\textit{It will make commits and diffs more readable, but might discard output (plots, etc.) you may actually want to store}'' [G1].

The next three best practices deal with the replicability of the execution environment in which notebooks are run. First of all, practitioners should always ``\textit{declare the dependencies in requirement files and pin the versions of all packages}'' [W2] they use within a project. An even stricter recommendation is shared in [G2], according to which ``\textit{all dependencies should be installed by the notebook itself}.'' For the same reason, \cBlue{five} articles recommend putting all import statements at the beginning of a notebook. As detailed in [G7], this practice ``\textit{has two benefits. The dependencies and tools used are obvious at the first glance. When you restart the notebook server, you can have all your imports restored with a single re-run. It is especially useful when you don’t want to re-execute the entire notebook.}''

To make the execution environment of computational notebooks even easier to reproduce, notebooks should be executed in self-contained environments. At the very least, virtual environments can be employed via environment managers like \texttt{conda}, \texttt{pipenv}, or \texttt{virtualenv} [G7]. In more complex scenarios, the full reproducibility of environments might be rather achieved by resorting to containers [W1, G2\cBlue{, G13}], e.g., Docker containers.

Even when notebooks are properly versioned and project dependencies are managed, the non-linear execution of code cells might produce hidden states that prevent the reproducibility of a computation. For this reason, \cBlue{six} articles suggest re-executing notebooks top to bottom before storing or sharing them with colleagues, e.g.: ``\textit{Many notebooks have out-of-order cells and skips. [...] We recommend re-running notebooks for restoring the execution counters and minimizing the impact of hidden states and out-of-order cells}'' [W2].

Also, W1 recommends regularly restarting the kernel of a notebook before its complete re-execution. It then goes even further suggesting to ``\textit{ask lab-mates or colleagues to try to run one of your notebooks and then listen when they explain any difficulties. Try to run their notebooks and let them know if you hit snags.}''

All the best practices around the theme of notebook traceability and reproducibility are inevitably of paramount importance in collaborative contexts.
Nowadays, it is unimaginable to develop software collaboratively without relying on version control systems \cite{brosch2009we}. On the other hand, being able to fully reproduce the computation performed by a colleague is necessary when a notebook is developed collaboratively, but also to ensure its future maintainability.

\textbf{Write high-quality code.}
Writing high-quality code is always important, and even more so when working in teams. 
As such, all the best practices related to code quality are deemed relevant for collaboration with computational notebooks.
Around this theme, we gathered best practices that mainly address code quality issues. 

\cBlue{One of the} best practice\cBlue{s} with the largest support (\cBlue{seven} articles, two from white literature and \cBlue{five} from grey literature), suggests modularizing code, i.e., to ``\textit{abstract code into functions, classes and modules}'' [W2].
This is in line with the recommendation found in [G1] for ``\textit{acquiring the habit of turning stable code into functions and move them to a dedicated module. This makes your notebook more readable and is incredibly helpful when productionizing your workflow}.''
\cBlue{
The right point in time in which notebook code should be modularized and refactored into a Python script is not trivial to determine; one should find a good compromise between the risk of technical debt \cite{kruchten_technical_2012} and that of premature optimization, as explained in [G11]:

\begin{quote}
    ``I also have a \textit{threshold of modularization}, where I determine if it’s better to refactor my Notebook into a Python script or module. On my end, once I see that my workloads become production- and cloud-heavy, then I make the effort to convert my notebooks into a Python script. The goal is to reduce tech debt and increase collaboration. [\dots] In addition, the threshold of modularization also reduces the risk for premature optimization (writing scripts early on in the project that may only be used once) and underengineering (using non-maintainable and clunky code to support mission-critical workloads).''
\end{quote}
}

Another theme that emerged is the importance of testing notebooks: \cBlue{five} articles [W2, G2, G3, G7\cBlue{, G9}] also mention testing as a best practice for working with computational notebooks. 


In software development, sticking to coding standards is a common way to ensure that the code produced by one developer will be easy to read, understand, and maintain by other developers as well \cite{boogerd_assessing_2008}. 
\cBlue{Four} articles [G1, G3, G5\cBlue{, G10}] suggest sticking to widely accepted coding standards (like PEP~8 in the case of Python) also when developing computational notebooks because ``\textit{having a standard in place will allow you to navigate easily between various projects of yours, as well as reduce the cognitive load about trying to decide what style to use here and there}'' [G3].

Giving notebooks meaningful names and using relative paths when referencing data files within notebooks are two further recommendations that we relate to the high-quality code theme. 
Their benefits are easily explained: a proper notebook naming strategy helps to navigate filesystems or shared repositories and finding the right notebook when needed; using relative paths, on the other hand, is an essential practice when notebooks need to be re-run on different machines. 
These might be thought of as basic coding standards, inspired by common sense. But common sense is not always common practice.

\textbf{Leverage the literate programming paradigm.} The success of computational notebooks among data scientists is largely due to the approach they enable to code documentation, inspired by the literate programming \cite{knuth_literate_1984} paradigm. 
The final result of a well-written notebook is a `book' describing the steps of computation and their rationale; this documentation, developed side by side with code, can be easily followed by its original author to reconstruct his work in the future, but it is especially useful when the notebook is shared. 
Collaborators can follow the reasoning behind analytical steps and straightforwardly intervene to fix or extend them. 
Moreover, well-documented notebooks can be used to transparently disseminate analytical results, enabling different kinds of stakeholders to examine the path that led to them.

\cBlue{Eight} articles [W1, G1, G3, G5, G8\cBlue{, G9, G10, G11}] contain best practices related to analysis documentation and \cBlue{six of them} [W1, W2, G1, G3\cBlue{, G9, G10}] recommend leveraging markdown headings to give computational narratives a clear and well-defined structure: \textit{``Document the process, not just the results. [...] make sure to document all your explorations, even (or perhaps especially) those that led to dead ends''} [W1]. 

\textbf{Keep your notebook clean and concise.} 
All the benefits of literate programming lose their strength if notebooks are messy or overly complicated. Nevertheless, during experimentation, notebooks can easily get untidy and thus difficult to read and understand \cite{head_managing_2019, rule_aiding_2018}. 
To address this issue, three articles [W1, W2, G3] report cleaning computational notebooks as a significant best practice. 
In particular, [W1] recommends to \textit{``clean, organize, and annotate your notebook after each experiment or meaningful chunk of work [...]''}, while [W2] -- more specifically -- warns to \textit{``pay attention to the bottom of the notebook''}, where typically unexecuted and poorly documented cells tend to stack up. 
More generically, [G3] recommends to beware of dead code in notebooks because of the ``\textit{added mental overhead}'' for data scientists.


Moreover, \cBlue{four} articles [W1, G5\cBlue{, G9, G10}] suggest keeping cells concise, e.g., by trying to make each of them \textit{``perform one meaningful step of the analysis''} [W1].
Also, the overall length of notebooks should be limited, splitting long ones into a series of notebooks referenced in a top-level index notebook [W1].

\textbf{Distinguish production and development artifacts.} \cBlue{Five} articles  [G2, G5, G6\cBlue{, G9, G11}] report an interesting best practice addressing the role of computational notebooks in a data science project with multiple contributors: distinguish between development artifacts (i.e., \textit{`lab'} or \textit{`dev'} notebooks) and production artifacts (i.e., \textit{`deliverable'} or \textit{`report'} notebooks). 
As defined in [G5]: \textit{``The difference between the two [...] is the fact that individuals control the lab notebook, while the deliverable notebook is controlled by the whole data science team.''}
A deeper characterization can be found in [G6]: the lab notebook \textit{``is not meant to be anything other than a place for experimentation and development''} and \textit{``keeps a historical (and dated) record of the analysis as it's being explored;''} on the other hand, deliverable notebooks are \textit{``fully polished versions of the lab notebooks''} storing the final output of the analyses.

\textbf{Embrace open dissemination.} The theme of open dissemination does not apply to all contexts. In typical industrial settings, analytical scripts and notebooks are not meant to be openly shared. 
However, notebooks are also employed in academic contexts in which, on the contrary, sharing replication packages that enable full analysis reproducibility is considered a virtue.
In any case, because they suggest better ways to share notebooks and data, these best practices are clearly related to collaboration and, therefore, are included in our catalog.

In particular, [W1] suggests making notebooks available by sharing them in public repositories with a clear \texttt{README} file and a liberal open-source license, granting permission for reuse; the recommendation goes even further, as the article suggests to share static and dynamic versions of notebooks that can respectively be read and executed requiring no local Jupyter installation.
The same article also advocates for data availability: \textit{``Strive to make your data or a sample of your data publicly available along with the notebook''} [W1] \cBlue{Another couple of articles [G9, G11] propose concrete strategies to enable data availability (e.g., using a cloud data bucket [G9] or a data version control system [G11]), while another article recommends that --- regardless of the data dissemination strategy --- notebook authors should clearly state where data should be retrieved from [G13]}.

\section{Interviews}
\label{sec:qualitative_study}

To gain a deeper understanding of the best practices from our catalog and to further ensure they provide satisfactory coverage of the subject matter, we conducted a series of semi-structured interviews with practitioners.

\subsection{Method}

\subsubsection{Interview participants}

Leveraging our network of professional contacts, we sent e-mail invitations to those working on data science projects. 
Twenty-two of them agreed to join the study: some basic details about each interviewee are reported in Table~\ref{tab:interviewees}.

\begin{table}[tb]
\centering
\footnotesize
\begin{tabular}{@{}lllll@{}}
\toprule
\textbf{ID}   &   \textbf{Gender}   &   \textbf{Exp. as DS (y)}   &   \textbf{Exp. with notebooks (y)}   &   \textbf{Job Role} \\ \midrule
\filan      & M    & 2      & 1.5   & Data Scientist            \\
\fuma       & M    & 11     & 5     & Senior Research Scientist \\
\rocco      & M    & 1      & 2     & Data Scientist            \\
\plamen     & M    & 1.5    & 3     & Data Scientist            \\
\sosi       & M    & 1.5    & 2.5   & Data Scientist            \\
\legna      & M    & 2      & 5     & Data Scientist            \\
\bakke      & M    & 10     & 4     & Data Scientist            \\
\andre      & M    & 2.5    & 2.5   & Data Scientist            \\
\ciuff      & M    & 2.5    & 2     & Data Scientist            \\ 
\dani       & M    & 3      & 3     & Senior Analyst            \\
\piemo      & M    & 2      & 2     & Data Scientist            \\
\giulia     & F    & 2      & 2     & Data Scientist            \\
\matt       & M    & 3      & 3     & Data Scientist            \\
\doss       & F    & 2      & 2     & Data Scientist            \\
\dista      & M    & 2      & 3     & Data Scientist            \\
\sann       & M    & 3      & 2     & Data Scientist            \\
\gabri      & M    & 2      & 4     & Data Scientist            \\
\sagli      & M    & 3      & 3     & Software Engineer         \\
\cordi      & M    & 2      & 3.5   & Product Manager           \\
\lanott     & F    & 3      & 2     & Senior Research Scientist \\
\tess       & M    & 3      & 2     & Senior ML Engineer        \\
\consort    & M    & 2      & 3     & Data Scientist            \\
\bottomrule
\end{tabular}
\caption{Gender, experience as data scientists (in years), experience with computational notebooks (in years), and job role of the interviewees}
\label{tab:interviewees}
\vspace{-8mm}
\end{table}

At the time of the interviews, all subjects had at least one year of experience as data scientists (median=2.5 years); the group also included a couple of senior data scientists, having 10 and 11 years of experience. 
Moreover, all subjects declared to have used Jupyter Notebook for more than one year; the less experienced one had been using computational notebooks for 1 year and a half.
The interviewees were working for companies in varied business domains, from IT consulting, to e-commerce and logistics.

\subsubsection{Semi-structured Interviews}

We conducted all interviews between August and September 2020, by using videoconferencing applications; each interview lasted about an hour.

At the beginning of each session, we tried to establish mutual trust by ensuring interviewees that their identity would not be disclosed nor their answers shared with their supervisors. 
After that, we broadly introduced the theme of collaboration with computational notebooks and asked them to share their personal best practices, as well as those followed by their teams. Each interview was guided by presenting the broader, main themes that emerged from the MLR (see Table~\ref{tab:MLR-summary}). For example, we asked what solutions they and their team put in place to make sure that analyses are traceable and reproducible. To avoid any bias, during the interviews we did not share the catalog of best practices; instead, we asked for clarification or to speculate further only after they mentioned a related topic (e.g., testing notebooks). We also invited the interviewees to freely share any other best practices followed to collaborate with computational notebooks.

During the interviews, the interviewer (the first author) took notes and made an audio recording of each session. 
Afterward, we transcribed the audio recordings and coded the transcripts together. 
In particular, we performed a closed coding by using as codes the catalog of specific best practices reported in Table~\ref{tab:MLR-summary}. This way, we were able to assess whether and to what extent the best practices are followed, according to the experience of professional data scientists, and possibly identify others that had not emerged from the MLR. However, during the interviews, none of the participants mentioned a best practice that was not traceable with those already available in the catalog.

\subsection{Results}

In this paragraph, we report the results of our closed coding of the interview transcripts. Before going through each best practice by theme, we point out that three of our interviewees (\fuma, \sagli, \cordi) firmly remarked on the importance of having a shared set of validated best practices within a data science team:

\begin{quote}
    ``The definition of a well-validated set of best practices must be a priority. 
    At this time, it would really be useful for my team to think about a set of best practices and try to apply them consistently within our projects, to see if they actually bring improvements'' (\fuma).
\end{quote}

\sagli{} and \cordi{} also underlined the value of sharing best practices to fill the cultural gap between data scientists and software engineers working together in data science teams. 

\subsubsection{Make your analysis traceable and reproducible{\cBlueSimple: It comes with trade-offs}}

BP1 (\textit{Use version control}), \cBlue{one of the two best practices} with the highest support in the MLR, was explicitly mentioned by only one interviewee (\andre), who commented: \textit{``One of the things that we do when putting notebooks under version control is to strip out the output of every code cell... in this way, at least we're sure to get lightweight and readable diffs.''} 
Besides, other interviewees mentioned the use of code repositories for storing notebooks while talking of other best practices (\fuma, \plamen, \ciuff). 
This suggests that, for many professional data scientists, this basic best practice might be taken for granted.

Eight interviewees (\filan, \fuma, \plamen, \sosi, \legna, \bakke, \andre, \ciuff) shared observations that concern BP2, advocating for managing project dependencies: \textit{``[Managing dependencies] is something we often do. We frequently need to move notebooks back and forth to different machines, with different environments''} (\filan). 
\fuma{} also added: \textit{``It also depends on the number of the dependencies. If they are a couple, I would place them at the top of my notebook, for example, using the \texttt{!pip install} expression. 
Otherwise, I would create a requirements file.''} 
\legna{} underlined the importance of adopting this practice while working with others, although -- in such collaborative scenarios -- it can be non-trivial to make requirement files consistently evolve over time (\sosi). 
A little more controversial is the idea of pinning versions for all packages (\filan, \ciuff), as explicitly suggested in [W2]: \textit{``We often use very specialized libraries that are seldom present on our clients' machines in the specific version we use on our development workstations. 
Usually, we try to mitigate this problem by taking the first two numbers of the version of a module and dropping the third''} (\filan).

A couple of interviewees (\fuma, \plamen) shared comments related to BP3 (\textit{Use self-contained environments}). 
\plamen{} pointed out: \textit{``Usually we have an environment per repository rather than an environment per notebook [...] this is definitely a best practice. 
However, to be honest, we don’t always do it. 
Sometimes we just have one environment with all the dependencies in there and use it all the time.''}

Five interviewees (\filan, \rocco, \legna, \bakke, \ciuff) expressed comments in agreement with BP4 (\textit{Put imports at the beginning}). 
Nevertheless, a couple of them were conflicted (\filan, \rocco): \textit{``It is something that I like, but I realize that it comes from my experience in writing traditional scripts. 
Often in notebooks, I write code blocks that are semantically consistent and self-contained, like the code to train a specific kind of model. Related imports are sometimes better placed close to the block rather than at the beginning of the notebook. This way, if the block is deleted, import statements that are no longer needed get deleted too''} (\filan).

Six interviewees (\filan, \rocco, \plamen, \legna, \bakke, \ciuff) also mentioned the importance of re-running notebooks top to bottom to ensure their re-executability (BP5).
\filan{} reported: ``\textit{I often [re-run notebooks top to bottom]. 
It does not save me from whatever could go wrong, but it is helpful, especially if done in a clean environment. 
It is the final check, isn't it?}'' 
However, one of the interviewees was conflicted (\rocco), as in his opinion explorative notebooks might legitimately contain slightly unrelated micro-tasks that are non-linearly executed; in such cases, a top-to-bottom re-execution \textit{``might not be that reasonable.''} 
Moreover, \plamen{} and \ciuff{} pointed out that a final, full re-execution of notebooks is always a sensible thing to do, unless the computational complexity of the analyses is high, making it a too expensive task to accomplish within the time allotted to the project.

\subsubsection{Write high-quality code{\cBlueSimple: Some consolidated best practices are controversial with notebooks}}

Most of the controversy found in the interviews is related to best practice BP6 (\textit{Modularize your code}). 
With the only exception of \andre, who firmly advocated for this recommendation, a notable amount of skepticism emerged from the comments of the other interviewees.
\fuma{} is in favor of implementing code abstractions like functions and classes, but only \textit{``as long as they stay within the notebook.''} 
His opinion is that notebooks should be simple enough to stay self-contained and not rely on external local modules: \textit{``If the complexity of code in a notebook raises to the point that one or more external modules are required, like for example a \texttt{utils.py} module containing all the utility functions, maybe developers have to look somewhere else for a most suitable development tool.''} 
A similar opinion was shared by \plamen. 
\filan{} mentioned code modularization as a general software development best practice that also applies to the context of computational notebooks, although admitting that, in general, the agreement on this idea might depend on the style in which notebooks are used. 
For instance, notebooks can be used in an `explorative style,' in which the time between thinking, coding, and getting results should be minimized as far as possible. 
In such a scenario, \textit{``this behavior might not be the one to follow, as it easily becomes counter-productive for the [practitioner's] goals.''}

As far as BP7 is concerned (\textit{Test your code}), \filan{} and \legna{} admitted that their usage is quite limited in notebooks.
\legna{} reported the amount of time assigned to a project to be a crucial factor in determining the feasibility of notebook code testing.
Besides, according to \filan's personal experience, in the best case, tests \textit{``come down to a series of assertions, to be put at the end of code blocks, just to make sure that they do what they are intended to do and trigger explicit errors otherwise.''}
Yet, ``\textit{assertions are simply not enough sometimes... you have to execute the whole notebook yourself and see if and where it fails... something that doesn't scale at all}'' (\sann). Indeed, assertions are difficultly wired into automated build pipelines; on the other hand, even if unit testing libraries can be used within notebooks (e.g., \texttt{unittest} and \texttt{pytest}), notebook implementations do not offer IDE-like facilities to perform tests and evaluate test coverage.
Accordingly, \bakke{} underlined that notebook verification is usually trickier than script verification: \textit{``What you can do with Python modules is to run them through \texttt{pytest}. I wouldn't know how to do that, for example, for a notebook cell.''} 
At odds with the other interviewees, \rocco{} and \ciuff{} argued that testing should not happen while working in a notebook environment, but only after a full transition to a traditional codebase.

Another best practice that received contrasting comments is BP8 (\textit{Name your notebooks consistently}). Albeit somewhat at odds with what we found in the literature [W2], some of the interviewees advocated for long filenames (\filan, \plamen, \sosi, and \ciuff) as, sometimes, they help to understand the content of notebooks at first glance.
\legna{} shared his habit of prefixing filenames with dates (in the case of  explorative notebooks) and sequential numbers (in the case of production notebooks); a sequential numbering of notebooks was also a best practice followed by \fuma. 
Furthermore, \bakke{} and his colleagues typically  specify parameter choices for ML experiments right in notebook filenames; this way, they can see what model the output of each notebook is by simply looking at its name.

Noticeably, only two interviewees mentioned BP9 (\textit{Stick to coding standards}). 
\filan{} pointed out that it is important to have a coherent coding style within a team, and that linters could help to achieve that.
\fuma{} reported that, at the time of the interview, his team was indeed trying to include code quality checks for notebooks -- based on \texttt{flake8}\footnote{A tool for Python style guide enforcement, \url{https://flake8.pycqa.org/en/latest}} -- in a continuous integration pipeline powered by TravisCI.\footnote{\url{https://travis-ci.org}}

Regarding BP10 (\textit{Use relative paths}), we recorded five favorable opinions (\fuma, \sosi, \legna, \andre, \ciuff). 
\fuma{} said: \textit{``I worked with many people who use absolute paths, referencing locations on their own file systems... I would use relative paths instead and place the referenced files in the same repository where the notebooks themselves live. Unless there is a company policy requiring a different behavior.''} 
\legna{} and \ciuff{} pointed out that achieving consistency on path conventions in a collaborative environment can be difficult, especially when working with junior data scientists: \textit{``I often find myself in trouble when someone else's in charge of the data organization, because I may not agree with their choices in terms of the project structure. 
This is usually a matter of expertise: senior data scientists tend to have an optimal project structure already in mind. 
With juniors, however, I find it more difficult to use relative paths''}  (\legna).

\subsubsection{Leverage the literate programming paradigm{\cBlueSimple: Narratives are key enablers to collaboration}}

Another area in which we registered consistent opinions (\filan, \fuma, \sosi, \bakke, \ciuff) was  related to BP11 (\textit{Document your analysis}). 
\sosi{} shared: \textit{``Markdown cells are super-useful! 
To me, notebooks are not just programs. 
I use them to narrate what happens at a business level and describe the logical steps that I do.''} 
He also pointed out that, ideally, by consolidating code abstractions (i.e., functions and classes) into external modules, the code inside notebooks should progressively leave pace to the Markdown descriptions. 
In this way, \textit{``notebooks are more about narrating the experiment -- or whatever one is doing -- rather than showing the code.''}
Following this idea, \consort{} added that ``\textit{computational notebooks might play a different but no less important role at the end of a project, by becoming part of the final documentation}''.
Other than being a crucial enabler of collaboration, the `storytelling' capability of notebooks is also considered vital for the reproducibility of analyses: \textit{``If you write notebooks methodically, then you create a story of your computations. And maybe in the future, you won't be able to re-run a notebook directly, but you can still follow the story and reproduce your successful experiment''} (\filan).

Overall, as far as BP11 is concerned, we detected some contrast only concerning where practitioners expect to find documentation. 
\rocco{} expects to find most of the descriptive text at the beginning of a notebook, while \filan{} remarks on the importance of documenting in the final part of an experiment: \textit{``Often notebooks lack a descriptive closure. You should always complete your notebooks by telling what you learned or achieved... The computational notebook should be used almost as if it were a technical blog post.''}

As for BP12 (\textit{Leverage Markdown headings to structure your notebook}), although three interviewees (\plamen, \sosi, and \legna) explicitly mentioned using Markdown headings in their work, none of them shared any interesting consideration in this regard.

\subsubsection{Keep your notebook clean and concise{\cBlueSimple: Definitely, but a little clutter toward the end is tolerable}}

Opinions related to BP13 (\textit{Keep your notebook clean}) were surprisingly controversial. 
On the one hand, most of the interviewees pointed out the importance of keeping notebooks tidy. 
Some of them also noticed that the last part of the notebook is where physiologically dirt builds up, as it is the place where experimental code is usually tried out before being put into place (\filan, \sosi, \legna, \bakke). 
\plamen{} stated: \textit{``I always try to keep it tidy, it doesn't matter if it's the end or middle or beginning... I just remove useless cells, especially if I have to push notebooks to a [shared] repository.''}
\fuma{} shared a less rigid recommendation: \textit{``The worst thing is to find non-executed or faulty cells in the middle of notebooks. 
However, if you find them at the end, after being able to execute the whole notebook without problems... you can understand...''}
On the other hand, \legna{} highlighted: \textit{``It depends on what you're doing. 
If you're developing a task that is meant to be executed via the notebook, then your notebook should definitely be clean, top to bottom. 
Otherwise, if you're just exploring or studying a dataset... then, having a bit of clutter at the end is quite standard.''}

Only \bakke{} mentioned the theme covered by BP14 (\textit{Keep your notebook concise}): \textit{``In principle, it's good to have short cells, as short as possible.''} However, he also added that, in specific circumstances, this best practice \textit{``might not hold.}'' 
As an example, he mentioned having had some problems with short cells when using a managed notebook system to launch batch notebook executions via API calls.

\subsubsection{Distinguish production and development artifacts{\cBlueSimple: Eventually, consolidate notebook content in a structured codebase}}

BP15 (\textit{Distinguish production and development artifacts}) was also mentioned by just one interviewee -- \bakke -- who shared how challenging it is to adopt this best practice:

\begin{quote}
    ``[...] what I really want to find out is how we should treat a notebook. 
    Like... there's a notebook that you're developing and a notebook that's for production. 
    The notebook that you're developing probably is in a state that is not production-ready. 
    And at some point, you need to choose which chunks of code are to be brought in the production container and which are not. 
    So, you kind of get this extra cycle in your development process, which includes cleaning up your stuff... and I would be really interested in learning what could be the optimal way of doing this.
    
    I guess that's a little bit different from regular Python development because typically, in that case, you work on a script and you know whether it works or not. 
    You can run tests against it. 
    And you can deploy it as it is. With a notebook... that’s a little bit different.''
\end{quote}

Other interviewees (\filan, \sosi, \piemo, \sann) were not particularly favorable to the idea of having  `development' (or `production-level') notebooks. They agreed that ``\textit{notebooks should be dismissed at a certain point during a project and code should be transferred to a more reliable and structured codebase}'' (P4).
\filan{} reported one of his recent experiences in which, after experiments had been carried out using individual computational notebooks, useful code parts were refactored and consolidated in traditional Python modules. 
He in person, as a software engineer, was in charge of refactoring code, while the data scientists in his team had performed the preliminary work on disposable, lab notebooks.

\subsubsection{Embrace open dissemination{\cBlueSimple: Taken for granted}}

None of our interviewees shared any thoughts on this theme. 
This is probably because, in an industrial context, the availability of notebooks and data within a team is taken for granted. 
However, this might be less obvious in different contexts, such as in the open-source or the scientific community.

\section{Quantitative Empirical Study}
\label{sec:quantitative_study}
\cBlue{Having gained an understanding of how experienced data scientists working in industry perceive the identified best practices, in this section, we perform a quantitative analysis to assess the extent of their adoption in practice.
As the data source, we chose Kaggle because the platform --- consistently with the collaborative scenario --- allows us to filter users by their expertise levels and, therefore, mine only the notebooks written collaboratively by expert data scientists.
}

\subsection{Materials and Methods}

In this section, we describe the dataset and the methods used in the quantitative validation of the best practices.

\subsubsection{Kaggle dataset}

To perform our quantitative analysis, we created a dataset of computational notebooks from Kaggle. 
At the core of the platform are \textit{Kaggle competitions}, challenges in which various types of rewards are offered, from fame to money prizes, and even job positions.
Users with any level of expertise can join competitions; nonetheless, Kaggle defines a clear progression system\footnote{\url{https://www.kaggle.com/progression}} to track their growth. 
The Kaggle progression system is designed around four categories of expertise, in which users can advance independently: \textit{Competitions}, \textit{Notebooks}, \textit{Datasets}, and \textit{Discussion}. To advance, users need to earn \textit{medals}: competition medals are awarded for top competition results; dataset, notebook, and discussion medals are gained based on their `popularity,'  i.e., the number of `upvotes' received respectively by one's artifact or post. Three types of medals can be awarded: bronze, silver, and gold. A notebook is awarded a bronze medal when it reaches 5 upvotes, a silver medal for 20 upvotes and a gold medal for 50.\footnote{Not all votes count towards the assignment of medals: i.e., self-votes, votes by novices, and votes on old posts are ignored.} Users can achieve one out of five progressive performance tiers within each category of expertise, namely \textit{Novice}, \textit{Contributor}, \textit{Expert}, \textit{Master}, and \textit{Grandmaster}.
The Expert tier is reached when the user completes ``\textit{a significant body of work on Kaggle in one or more categories of expertise},'' as it requires the achievement of a minimum number of bronze medals that varies according to the category.

A large number of notebooks are publicly available on the platform, which also provides \textit{Meta Kaggle},\footnote{\url{https://www.kaggle.com/kaggle/meta-kaggle}} an official, daily updated collection of public metadata on users, competitions, datasets, and notebooks. 
Thanks to Meta Kaggle, which we downloaded in October 2020, we were able to select Jupyter notebooks with up-to-date language and author information.
We kept only the notebooks written in  Python and consequently discarded the few existing R notebooks (\textasciitilde$3\%$ of the full dataset).
Moreover, we leveraged Meta Kaggle to collect evidence of collaboration while working with the selected computational notebooks.

\subsubsection{Notebook selection criteria}

\begin{figure}
    \centering
    \includegraphics[width=\textwidth]{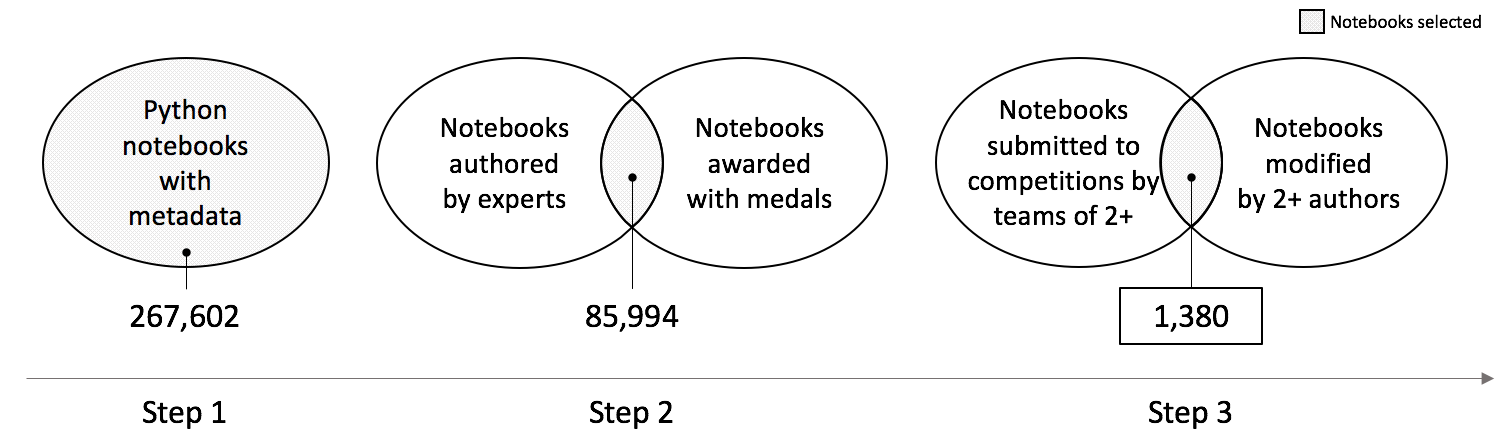}
    \caption{Filtering steps applied to select notebooks from Kaggle}
    \label{fig:notebook-criteria}
\end{figure}

We selected notebooks from Kaggle by applying the following filters in three consecutive steps (see Figure~\ref{fig:notebook-criteria}).

\textit{\textsc{Step 1}: selection of Python Jupyter notebooks with full metadata.} We identified all the Jupyter notebooks in Python for which Meta Kaggle provides metadata about their current kernel version, the programming language, and the author. We thus selected a total of \numprint{267,602} notebooks.

\textit{\textsc{Step 2}: selection of high-quality notebooks.} To select high-quality notebooks, we resorted to two different criteria: the first was to consider notebooks developed by experienced authors (\numprint{75,591} notebooks); accordingly, we followed the operational definition provided by Kaggle of an \textit{expert user}, defined as a user who has attained at least the `Expert' performance tier in at least one of the four `Categories of Expertise'.\footnote{Because Meta Kaggle does not contain information about the performance tier achieved by its users in the Datasets category of expertise, we resorted to the remaining three categories to determine the subset of experienced Kaggle users.}
The second criteria was to select notebooks that had been awarded at least one medal (\numprint{23,229} notebooks).
  
We joined the filtered sets of notebooks resulting from these criteria, removed duplicates, and used the resulting \numprint{85,994} notebooks as a base for further filtering.

\textit{\textsc{Step 3}: selection of collaboratively developed notebooks.} To select the collaboratively developed notebooks, we adopted another couple of criteria.
First, we considered notebooks whose last version had been submitted to a competition by a team comprising two or more users. This was not obvious because, in Kaggle, users join competitions as teams, but many comprise just one user.
By using this filtering criterion, we were able to retrieve \numprint{888} notebooks.
Second, we looked at the version history of Kaggle notebooks and selected those in which different versions had been authored by at least two distinct users. We thus identified \numprint{593} notebooks.
Finally, as done in step 2, we joined the two sets and dropped duplicate notebooks.

Upon performing the filtering steps described so far, we retrieved \numprint{1,386} Kaggle notebooks, of which 6 were dropped as duplicates (although having different names).
Therefore, we eventually built a dataset of \numprint{1,380} high-quality, valid Python notebooks created collaboratively between November 2015 and October 2020.

\subsubsection{Tools}

To perform the quantitative validation of the best practices from our catalog, we modified and extended the scripts from a previous study on the reproducibility of Jupyter notebooks in GitHub~\cite{pimentel_large-scale_2019}. 

Our complete replication package\footnote{\url{https://doi.org/10.5281/zenodo.4441236}} includes the scripts, the dataset, and an additional Python module that we developed to study the compliance of code contained in computational notebooks to the PEP~8 style guide specifications. The module is based on \texttt{Pylint},\footnote{\url{www.pylint.org}} a quality assurance tool for Python code which supports the following five categories of checks:
\begin{enumerate*}
\item \textit{convention}, to enforce conformance to PEP~8;
\item \textit{refactoring}, to suggest improvements to the structure of code;
\item \textit{warnings}, to notify possible problems in the code;
\item \textit{errors}, to notify possible bugs in the code;
\item \textit{fatal}, to log severe errors preventing further processing.
\end{enumerate*}

Our module operates as follows. First, it groups Jupyter notebooks by their Python version. 
Indeed, \texttt{Pylint} checks are specialized for the specific Python version of the target code (i.e., to lint a Python 2.7 script, \texttt{Pylint} has to be invoked from a Python 2.7 interpreter). 
In particular, we had to consider Python versions from 3.5 to 3.7, as they were the only ones used by the notebooks in our dataset. 

Then, our module runs \texttt{Pylint} against each group. In particular, for each batch of notebooks, it prepares a \texttt{conda}\footnote{\url{https://docs.conda.io/en/latest}} environment with a compatible Python version, pre-processes each notebook by extracting the source code from its code cells, and finally invokes \texttt{Pylint} on the extracted code, collecting the results in a \texttt{.csv} file.

During code extraction, our module carefully removes notebook-specific statements that might be detected as errors by \texttt{Pylint} (e.g., IPython magic commands \cite{perez_ipython_2007, vanderplas_jake_ipython_2016}).
Moreover, it uses a \texttt{Pylint} configuration set up to ignore checks that are prone to false-positive results in the context of computational notebook code (e.g., \textit{pointless-statement}).

\subsubsection{Operationalization of the best practices}

In this paragraph, we present the operational definitions that we assigned to the best practices in our catalog. 
Not all the best practices can be straightforwardly operationalized; moreover, because some operationalizations cannot be applied to the context of Kaggle, the related best practices have been excluded from our quantitative analysis.

The first three best practices (BP1, BP2, and BP3) could not be operationalized because, in Kaggle, the platform itself takes care of keeping notebooks under version control, and the Notebooks IDE --- Kaggle's implementation of Jupyter Notebook --- is served in the form of a self-contained environment (based on Docker containers), in which project dependencies are inherently managed.

To operationalize BP4 (\textit{Put imports at the beginning}), we broadly determined where import statements are located within notebooks. 
\cBlue{We decided not to consider the `beginning of the notebook' in its strictest sense, i.e., limited to the first code cell, as such cell may be used for other configuration purposes (e.g., to define environment variables or to install requirements via the \texttt{!pip install} expression)}.
Accordingly, after processing the whole dataset, we generated a plot that shows the percentage of code cells including import statements at the beginning, in the middle, and at the end of the sampled notebooks.

To assess the degree of adoption of BP5 (\textit{Ensure re-executability}), we considered the number and rate of notebooks presenting a sequential execution order of code cells --- starting from 1 --- without skips or repeated executions.
\cBlue{Indeed, that is the only possible state in which a notebook might be left after a final top-down re-execution.}

To evaluate the adoption of BP6 (\textit{Modularize your code}), we calculated the number and rate of notebooks including (1) local module imports, (2) function definitions, and (3) class definitions.

Similarly, we operationalized BP7 (\textit{Test your code}) by considering the number and rate of notebooks employing testing libraries. 
To identify them, we looked for imported modules with names containing the words `test,' `Test,' `TEST,' `mock,' Mock,' or `MOCK' as a sub-string, as well as other well-known Python testing packages not containing the same sub-strings (e.g., Nose2\footnote{\url{https://docs.nose2.io/en/latest}} and Robot).\footnote{\url{https://robotframework.org}}
\cBlue{Following the same approach adopted by~\fullcite{pimentel_large-scale_2019}, we did not include assertions in the operationalization of BP7 because they can be used for a broad range of purposes not necessarily related to code testing (e.g., verify data properties and statistical assumptions).}

BP8 (\textit{Name your notebooks consistently}) summarizes three different recommendations on notebook naming, the most basic of which is ``Don't forget to name your notebook documents!'' [G5]. 
In Kaggle, this cannot happen: even if a user forgets to give their notebook a customized name, the platform assigns a name automatically, by appending a hash code to the word `notebook.' 
Moreover, the platform enforces the use of a compatible character set. 
In the end, giving notebooks meaningful names, possibly sticking to a set of rules or conventions, is considered useful to easily recognize notebooks when browsing a filesystem or a project repository [G5, G6]. However, this is not the case when working in Kaggle, therefore we did not operationalize this best practice.

We evaluated BP9 (\textit{Stick to coding standards}) by employing our linting module for the assessment of computational notebook code compliance to the PEP~8 coding standard.
\cBlue{Specifically, we based our linting module on the Pylint library because, other than analyzing code style, it also checks for possible code errors and provides useful suggestions for refactoring.
Accordingly, the} operationalization \cBlue{of BP9} included the percentage of notebooks that failed at least one check for each of the \cBlue{available} Pylint check categories: `Convention,` `Warning,' `Error,' and `Refactor.'

BP10 (\textit{Use relative paths}), could not be properly operationalized in the context of Kaggle. 
Indeed, when a Kaggle user adds a dataset to their notebook, the platform mounts the corresponding directory at a fixed path within the underlying computational environment (i.e., \texttt{/kaggle/input}). Also, each new notebook is executed in an environment that has the same standardized filesystem structure and, when a user opens a shared or forked notebook within the platform, the notebook gets executed in an exact replication of its original environment, with the same filesystem. 
As a result, absolute paths in Kaggle are not as harmful as they are outside the platform, and thus their usage cannot be considered a bad practice.

As for BP11 (\textit{\textit{Document your analysis}}), we estimated the extent to which data scientists comply with this best practice by studying the presence of Markdown (MD) text in their computational notebooks. 
First of all, we leveraged a similar plot to that used in the operationalization of BP4, showing the distribution of MD cells compared to code cells within notebooks.
Then, we calculated the number and rate of notebooks containing MD text. 
We also computed the 5-number summary \cBlue{(i.e., the minimum, the three quartiles, and the maximum)} of the number of MD cells and code cells, as well as the number of meaningful MD words and lines\footnote{By `meaningful MD words' we mean all the words that are not part of the MD syntax. Accordingly, by `meaningful MD lines' we mean lines that have meaningful words.} in notebooks.

BP12 could be operationalized only partially, as there is no way to determine whether MD headings actually contribute to give notebooks a good structure. We therefore only assessed the number and rate of notebooks using MD headings and computed a 5-number summary of MD headings length.

To operationalize BP13 (\textit{Keep your notebook clean}), we assessed how executed, non-executed, and empty code cells are distributed in notebooks, as we considered them as evidence of untidiness. 
Although we understand that cleaning notebook code potentially involves more complex operations than removing useless cells, assessing the cleanliness of Python code, in general, is outside of the scope of this paper.

To estimate notebooks conciseness (BP14), we computed the 5-number summary of the number of cells per notebook, the number of lines and Python lines per notebook, and the number of lines in cells, code cells, and MD cells in notebooks.

Finally, the last three best practices (BP15, BP16, and BP17) could not be operationalized. As far as BP15 is concerned (\textit{Distinguish production and development artifacts}), there does not seem to be such a thing as a `production' stage within the Kaggle platform. 
Similarly, we had no way to assess the rate of adoption of BP16 (\textit{Make your notebooks available}) and BP17 (\textit{Make your data available}), which are inherently reinforced in Kaggle.

\subsubsection{\cBlue{Manual coding}}

\cBlue{For some of the guidelines, we found that a thorough evaluation of their adoption in notebooks required human judgment. 
For instance, the numeric measures computed in our quantitative analysis suggest broad trends in notebook documentation practices, but it is hard to tell whether a notebook is well documented by merely looking at the number of its MD cells or lines. 
For this reason, we decided to complement our quantitative analysis by performing a manual analysis of the notebooks for the four best practices deemed more subjective and controversial, namely: BP11 (\textit{Document your analysis}), BP12 (\textit{Leverage Markdown headings to structure your notebook}), BP13 (\textit{Keep your notebook clean}), and BP14 (\textit{Keep your notebook concise}).

We started by randomly selecting a representative sample of 301 items from our dataset of 1,380 Kaggle notebooks.\footnote{The size of a representative sample of the complete corpus with a 5\% margin of error and 95\% confidence level.} 
Then, the first two authors of the paper independently engaged in a pilot, open-coding session, after which they were able to define a closed set of codes. 
Subsequently, after independently coding the first 90 notebooks,  the inter-rater reliability was evaluated by computing the Cohen's $\kappa$ coefficient, which showed a \textit{substantial} (.78) to \textit{strong} (.87) agreement across the four best practices. 
Finally, after reconciling the identified discrepancies, the first author completed the coding of the remaining notebooks. 
More details on the process are outlined in Appendix~B. 
The results are reported in the following section, as a complement to the related findings obtained from the quantitative analysis.}

\def \firstcolwidth {2.2cm}
\def \secondcolwidth {2.3cm}

\begin{table}[h!]
\centering
{\scriptsize
\def\arraystretch{1.5}
\begin{tabular}{@{}p{\firstcolwidth}p{\secondcolwidth}p{4.5cm}p{4cm}@{}}

\toprule

\textbf{Theme} & \textbf{Best practice} & \textbf{Operationalization} & \textbf{Result}\newline5-number summaries are reported as\newline [min, q1, median, q2, max] \\

\midrule

\multirow{2}{\firstcolwidth}{Make your analysis traceable and reproducible} 
 & BP4 \textit{Put imports}\newline\hphantom{BP4}\textit{at the beginning} 
 & Distribution of import~statements~\textcolor{table_blue}{$\blacksquare$} in notebooks
 & \raisebox{-.7\height}{\includegraphics[width=4cm]{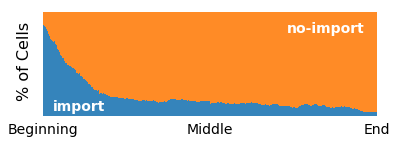}} \\
 
 \cmidrule(l){2-4}
 
 & BP5 \textit{Ensure}\newline\hphantom{BP5}\textit{notebook}\newline\hphantom{BP5}\textit{re-executability} & Notebooks executed top to bottom & 1,103 (94.35\%) \\
 
 \midrule

\multirow{8}{\firstcolwidth}{Write high-quality code} & \multirow{3}{\secondcolwidth}{BP6 \textit{Modularize}\newline\hphantom{BP6} \textit{your code}} & Notebooks with local module imports & 8 (0.59\%) \\
 &  & Notebooks with function definitions & 1006 (74.35\%) \\
 &  & Notebooks with class definitions & 343 (25.35\%) \\
 
 \cmidrule(l){2-4}
 
 & BP7 \textit{Test your code} & Notebooks with test modules & 1 (0.07\%) \\
 
 \cmidrule(l){2-4}
 
 & \multirow{4}{\secondcolwidth}{BP9 \textit{Stick to coding}\newline\hphantom{BP9}\textit{standards}} & Notebooks with failing `Convention' checks & 74.49\% \\
 &  & Notebooks with failing  `Warning' checks & 68.39\% \\
 &  & Notebooks with failing `Error' checks & 55.74\% \\
 &  & Notebooks with failing `Refactor' checks & 29.58\% \\
 
 \midrule
 
\multirow{8}{\firstcolwidth}{Leverage the literate programming paradigm} & \multirow{6}{\secondcolwidth}{BP11 \textit{Document}\newline\hphantom{BP11} \textit{your analysis}} & Distribution of markdown~cells~\textcolor{table_blue}{$\blacksquare$} and code~cells~\textcolor{table_orange}{$\blacksquare$} in notebooks
 & \raisebox{-.7\height}{\includegraphics[width=4cm]{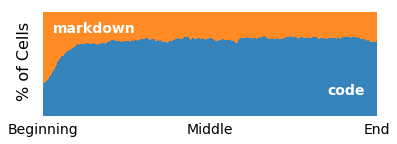}} \\ \\
 &  & Notebooks with markdown & 1,147 (83.12\%) \\
 &  & Code cells in notebooks & [0, 11, 19, 30, 205] \\
 &  & MD cells in notebooks & [0, 2, 7, 14.25, 220] \\
 &  & Meaningful MD words & [0, 60, 231, 606, 10971] \\
 &  & Meaningful MD lines & [0, 13, 30, 57, 1316] \\
 
 \cmidrule(l){2-4}
 
 & \multirow{2}{\secondcolwidth}{BP12 \textit{Use MD headings}\newline\hphantom{BP12} \textit{to structure}\newline\hphantom{BP12} \textit{your notebook}} & Notebooks with MD headers & 1,014 (88.40\%) \\
 &  & Median MD header size (words) & [0, 7, 19, 40, 1223] \\
 
 \midrule
 
\multirow{8}{\firstcolwidth}{Keep your notebook clean and concise}

 & \multirow{2}{\secondcolwidth}{BP13 \textit{Keep your}\newline\hphantom{BP13} \textit{notebook clean}} & Distribution of executed~\textcolor{table_blue}{$\blacksquare$}, non-executed~\textcolor{table_orange}{$\blacksquare$},\newline and empty~\textcolor{table_green}{$\blacksquare$} cells in notebooks 
 & \raisebox{-.7\height}{\includegraphics[width=4cm]{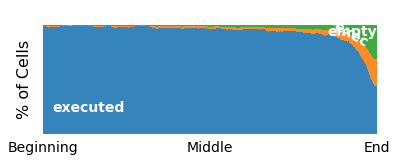}} \\
 &  & Empty cells in notebooks & [0, 0, 0, 1, 33] \\
 
 \cmidrule(l){2-4}
 
 & \multirow{6}{\secondcolwidth}{BP14 \textit{Keep your}\newline\hphantom{BP14} \textit{notebook concise}} & Number of cells per notebook & [1, 16, 26, 43, 365] \\
 &  & Number of lines in cells & [1, 1, 3, 9, 5643] \\
 &  & Number of lines in code cells & [1, 1, 4, 12, 5643] \\
 &  & Number of lines in markdown cells & [1, 1, 1, 3, 259] \\
 &  & Number of lines in notebooks & [6, 133, 248, 440.25, 5995] \\
 &  & Number of Python lines in notebooks & [1, 103, 222, 399, 5981] \\ \bottomrule
\end{tabular}
}
\caption{Summary of the results from the quantitative validation of our best practice catalog}
\label{tab:quantitative-results}
\end{table}

\subsection{Results}

In this paragraph, we report the results of the quantitative validation of the best practices from our catalog, summarized in Table~\ref{tab:quantitative-results}.

\subsubsection{Make your analysis traceable and reproducible{\cBlueSimple: The effect of enforcing a commit policy}}

BP4 seems to be adopted in the majority of the notebooks in our dataset, as revealed by the plot on the distribution of import statements within notebooks. 
Indeed, the percentage of cells containing import statements rapidly decreases immediately after the notebook beginning and then keeps diminishing at a slower pace until the final part of the document.

As regards BP5, \numprint{1,103} out of \numprint{1,169} executed notebooks\footnote{By `executed notebooks' we mean notebooks that include the output of code cells and their execution counters. \numprint{1,169} of such notebooks were found in our dataset (84.71\% of the entire sample).} (94.35\%) present ordered execution counters without skips, a piece of evidence that they have been re-executed before being committed.
This very high rate can be explained by the fact that Kaggle reinforces this good practice: every time a user wants to submit the results to join a competition, they have to click the \textit{Commit (Save \& Run All)} button, thus verifying and consolidating the notebook as recommended in the literature.

\subsubsection{Write high-quality notebooks{\cBlueSimple: Code abstractions are common,  testing is not, and code quality is poor}}

Regarding the modularization of code (BP6), almost three-quarters of the notebooks (74.35\%) present at least one function definition, while slightly more than one quarter (25.35\%) contains a class definition. 
Local module imports appear only in 8 notebooks, accounting for 0.59\% of the entire dataset. This can be partly explained by the fact that Kaggle notebooks are conceived as standalone artifacts that are not stored in repositories together with other project files. Nonetheless, Kaggle users can create utility modules and import them into their notebooks.\footnote{\url{www.kaggle.com/product-feedback/91185}} 
Despite the availability of this feature, announced in 2019, the use of local modules appears to be still a rare exception in Kaggle.

As for BP7, even rarer is the use of testing modules: we were able to find a test-related import (\texttt{pytest} package) in only one notebook across the entire collection  (0.07\%).
A remarkably high percentage of notebooks (74.49\%) violates the checks related to the coding conventions (BP9). 
The most common violations are \texttt{bad-whitespace} (\numprint{51,693} hits), a check including all coding conventions related to the use of whitespace in code,\footnote{Examples of more fine-grained related messages are: \textit{``No space allowed before comma''} and	\textit{``Exactly one space required after comparison.''}} \texttt{invalid-name} (\numprint{41,024} hits), a check triggered when a name does not fit the naming convention associated with its type (constant, variable, class, etc.), and \texttt{trailing-whitespace} (\numprint{17,804} hits).

Similarly, we found a high number of notebooks for which \texttt{Pylint} raised warnings (68.39\%). 
The most common are the  \texttt{redefined-outer-name} warning, raised when the name of a variable hides a name defined in the outer scope, and the import-related warnings (\texttt{unused-import} and \texttt{unused-wildcard-import}).

More than half of the notebooks failed `Error' checks (55.74\%). 
The most frequent is \texttt{undefined-va}-\texttt{riable} (\numprint{1,795} hits), followed by \texttt{no-name-in-module} (348 hits), and \texttt{syntax-error} (339 hits).

Finally, the failed checks in the `Refactor' category are less common than the others: they are issued for 29.58\% of the notebooks. 
The most common violations are \texttt{too-many-locals} (422 hits), issued when a method or function uses too many local variables, \texttt{too-many-arguments} (399 hits), concerning the high number of arguments in function definitions, and \texttt{no-else-return} (268 hits), indicating the presence of unnecessary \texttt{else} statements after \texttt{if} statements containing a \texttt{return}.

\subsubsection{Leverage the Literate Programming paradigm{\cBlueSimple: Experts do use Markdown cells, but documentation quality is low overall}}

BP11 suggests always documenting the analysis in computational notebooks and appears to be adopted to a good extent in our dataset. 
The graph shows that the distribution of the MD cells is quite even within the analyzed notebooks, exception made for the beginning when they tend to be even more frequent. 
Moreover, we found that 83.12\% of notebooks in our sample contain MD text. The median number of code cells in notebooks is 19, only a few units higher than the median number of MD cells. 
Hence, the ratio between code and documentation text is rather balanced. 
Besides, we found that notebooks have a median of 30 meaningful MD lines and 231 meaningful MD words.
\cBlue{Notwithstanding these encouraging quantitative results, the manual inspection revealed that 64.63\% of the notebooks appear to be poorly documented and that in 20.73\% of them the documentation is only kept to a decent minimum.}

We found that \cBlue{MD headings} are used in the vast majority of \cBlue{the notebooks} (88.40\%), with a median length of 19 words. 
\cBlue{In this case, the manual coding confirmed that, overall, the use of MD titles is satisfactory (judged as \textit{Good} in 53.66\% of the notebooks and  \textit{Fair} in 7.32\% of them). 
However, there is still large room for improvement, as in 39.02\% of the notebooks the headings do not contribute to give the document a proper structure.}

\subsubsection{Keep your notebook clean and concise{\cBlueSimple: Experts comply}}

Regarding BP13, the plot in Table~\ref{tab:quantitative-results} shows that the majority of notebooks can be considered `clean' as  the overall number of empty and non-executed cells is quite low; 
moreover, they tend to accumulate towards the end of notebooks. 
The median number of empty cells in notebooks is 0.
\cBlue{Even more encouraging are the results from the manual coding, which revealed that 80.49\% of the notebooks are perfectly clean.}

Regarding conciseness (BP14), the notebooks in our dataset have a median of 26 cells, and 248 lines (222 lines of Python code). Overall, cells are composed of 3 lines (4 lines of code and 1 line of MD in median). 
\cBlue{The manual inspection revealed that, in most of the notebooks, the adherence to this best practice is mostly \textit{Good} (59.76\%) or \textit{Fair} (21.95\%)}.

\subsection{\cBlue{Best Practices in The Most Upvoted Notebooks}}

\cBlue{To complete our quantitative analysis, we decided to assess whether best practice compliance increases with the overall quality of Jupyter notebooks.
While the analyzed dataset of high-quality notebooks has been selected from Kaggle based on the expertise of the respective authors, the platform also allows estimating the popularity of a kernel through the number of received upvotes, which might be considered a further hint of notebook quality. 
Accordingly, we took upvotes into account in our experiment by repeating the quantitative analysis on two subsets of the original dataset (for brevity, hereafter referred to as \textit{ALL}): the first, \textit{P75}, containing the  348 notebooks in the 75th percentile (upper quartile) of the original dataset ordered by the number of upvotes; the second \textit{P90} comprising the 140 notebooks in the 90th percentile.

Albeit not large, the observed differences confirm that the most upvoted notebooks are in general more compliant with the identified set of best practices. For instance, we found fewer notebooks that fail \texttt{Pylint} checks (e.g., the relative frequency of notebooks failing `Convention` checks was 74.49\% in \textit{ALL}, 71.39\% in \textit{P75}, and 72.86\% in \textit{P90}, while the relative frequency of notebooks failing `Warning' checks was 68.39\% in \textit{ALL}, 57.8\% in \textit{P75}, and 45\% in \textit{P90}); the number of notebooks re-executed top to bottom is slightly higher in \textit{P75} (97.30\%) and \textit{P90} (96.80\%) than in \textit{ALL} (94.35\%); also, the rate of notebooks containing MD text is higher in the two subsets (\textit{ALL}=83.12\% vs. \textit{P75}=88.79\%, \textit{P90}=92.14\%); consistently, the median numbers of MD cells, meaningful MD lines, and meaningful MD words are higher in \textit{P75} and \textit{P90}, whereas the median number of code cells is lower. 
The rate of notebooks leveraging MD headings is higher in the most upvoted notebooks (\textit{P75}=92.23\%, \textit{P90}=95.35\%) than in \textit{ALL} (88.40\%). 
However, we also found some of the measures to remain almost constant across the datasets (e.g., the distribution of import statements; the distribution of non-executed and empty cells; the median number of empty cells; the rate of notebooks using testing modules or importing local ones). 
Instead, the use of function definitions within notebooks is less common in the most upvoted notebooks (\textit{P75}=73.39\%, \textit{P90}=58.39\%) than in \textit{ALL} (74.35\%);

Overall, despite a slight tendency for the most upvoted notebooks to be more compliant with the best practices, the small differences observed in the three samples are arguably explained by the fact that the original dataset in our study already comprises high-quality notebooks, as they are filtered according to authors’ expertise; indeed, getting notebook upvotes is one of the possible ways to progress through the Kaggle Progression System. 

For a more detailed comparison of the measures observed for the three datasets, please refer to the summary table stored in the replication package of this study on Zenodo.
}

\section{Discussion}
\label{sec:discussion}

We reviewed both white and grey literature and built a catalog of 17 best practices for collaboration with computational notebooks,  
albeit without any empirical validation (e.g., through a field study). 
Almost all the best practices from our catalog (15 out of 17) were mentioned by our interviewees as best practices that they follow (or, at least, strive to follow) when working collaboratively in their teams. Further encouraging results come from our quantitative study, in which most of the best practices that could be operationalized appear to be adopted by experienced data scientists on Kaggle. However, for some of the recommendations, the opinions of our interviewees were somewhat divided; in a few cases, even the findings from our quantitative study have shown that their adoption rate is far from ideal. 

\textbf{Explanatory versus exploratory notebooks}. Overall, there seems to be a conflict between what experts believe to be ideal and what they find reasonable to do when they need to be pragmatic at work.
This is why the most neglected best practices tend to be costly to implement (code modularization, testing, development of standards-compliant code).
Echoing the words of one of our interviewees, when notebooks are used for exploration, ``\textit{the time between thinking, coding, and getting results should be minimized as far as possible}'' (\filan). Code quality trade-offs in exploratory programming are not new to the literature \cite{beth_kery_exploring_2017, brandt_opportunistic_2008,kery_variolite_2017}. Rule et al.~\cite{rule_exploration_2018} showed the existence of a tension between `exploration' and `explanation' in computational notebooks, arguing that the exploratory process of data analysis tends to be a source of messes (e.g., alternative code, duplicate cells, etc.) that hinder what is arguably the major potential of notebooks: the ability to present an analytical process and its results along with a clear explanation. 
\cBlue{Our findings provide further evidence} that experienced authors of Jupyter notebooks undergo a similar tension between the need for a rapid exploration/prototyping workflow and the production-readiness of their artifacts\cBlue{, as also put forward by \fullcite{wang_human-ai_2019} when observing the `scatter-and-gather' phases of collaboration in data science.}

\textit{Implication for computational notebook platform designers}. To promote high-quality notebooks, taking this tension into account, notebook development environments should implement features that ease the adoption of best practices. 
\cBlue{Some of them might be automatically ensured in a way that is fully transparent to the notebook end-user and does not have an impact on his workflow; an example is BP2 (\textit{Manage project dependencies}): the list of packages required to execute and reproduce a notebook might be shipped with the notebook itself (and consistently updated over time), thus becoming part of its metadata as in the case of the kernel Python version.
Other best practices might be encouraged with the help of unobtrusive and customizable recommendations, to be issued at specific times during a workflow (e.g., before committing the notebook to a shared repository); BP5 (\textit{Re-run notebooks top to bottom}) is agreed upon by most of the interviewees and has clear benefits on notebook reproducibility: by following the example of Kaggle, notebook development platforms might encourage it by letting users choose, at the end of each experimental session, whether to save a notebook as it is or to commit it after a complete and successful re-execution.}

\textbf{Recommendations may need to be adapted to your context}. BP4 (\textit{Put imports at the beginning}) is a best practice on which most of the interviewees agreed upon. Nonetheless, a couple of them were conflicted and advocated for the opposite behavior: putting imports only where needed within a notebook. The results from the quantitative study showed that experts tend to require packages at the beginning, but also that there is room for exceptions: indeed, import statements were found across the whole range of notebook positions. \cBlue{This is consistent with the results by \fullcite{pimentel_large-scale_2019} from their large-scale study of 
GitHub notebooks.} Similarly, the best practices recommending a top-to-bottom re-execution of notebooks (BP5) or to keep notebooks clean (BP13) were mostly favored by the interviewees, although with noteworthy exceptions. Summing up the results from the quantitative study, the same pattern emerged: the majority of experts from Kaggle seem to largely adopt both best practices \cBlue{--- even more than GitHub contributors, in the case of BP5\footnote{The rate of notebooks showing a top-to-bottom re-execution is 94.35\% in our dataset and 56.1\% in the dataset by Rule et al.} \cite{rule_exploration_2018}} --- but findings contain also evidence of different approaches. We speculate that recording diverging opinions regarding the recommendations in our catalog is somehow physiological, as their appropriateness might depend on contextual factors other than the expertise of a practitioner (e.g., the goal of the notebook, the collaboration dynamics, etc.).

\textit{Implication for researchers}. We encourage future work to validate the best practices for collaboration with computational notebooks by taking into due account the existence of contextual factors that might influence the suitability of a recommendation.

\textbf{Code modularization should  happen inside notebooks}. A different pattern emerged for BP6 (\textit{Modularize your code}), one of the most controversial recommendations in our interviews. For one of our interviewees, notebooks should be self-contained and abstractions are only good as long as they stay inside them. The priority is to get a full understanding of how a notebook works as quickly as possible. Any time spent exploring one or more local modules external to the notebook is counterproductive. This seems to be particularly true in Kaggle, where --- even if allowed by the platform --- the use of local modules (or `utility scripts') appears to be a rare exception;
\cBlue{interestingly, the rate of notebooks with local imports is only slightly higher in GitHub (10.30\%)~\cite{pimentel_large-scale_2019}.}
Other interviewees pointed out that putting this recommendation into practice with notebooks is challenging; one of them argued that, when notebooks are written in an `explorative style,' switching to a different environment to create abstractions or refactor your code is counterproductive.

\textit{Implication for computational notebook platform designers}. Computational notebooks should provide native support to developers for consolidating portions of code into external Python modules. One such example is the set of Jupyter Notebook extensions proposed by \fullcite{head_managing_2019}. \cBlue{To be unobtrusive and not interfere with data exploration activities and fast prototyping, refactoring-oriented extensions might be collected in a new UI perspective, integrated into notebook development environments; this `refactoring perspective' should be normally hidden, but ready to be activated by the user when they need to polish their work and make it ready for sharing or production.}

\textbf{Need for notebook-specific testing frameworks}. BP7 (\textit{Test your code}) resulted in another controversial theme: a couple of interviewees argued that testing should not happen while working in notebooks, as it should become a concern only after a full transition of the code to a traditional codebase. The quantitative results seem to confirm this view, or at least are consistent with it, as we found evidence of testing in only one notebook in our dataset.
\cBlue{This also confirms what has been observed in GitHub, where only 1.54\% of the analyzed notebooks included import statements for Python testing modules~\cite{pimentel_large-scale_2019}.}
Nevertheless, the remaining interviewees recognized the importance of testing code, although remarking various obstacles, i.e., the lack of time and tool support for testing within notebooks.

\textit{Implication for computational notebook platform designers}. Our findings suggest an opportunity to develop a lightweight testing method capable of supporting the fast and iterative usage of notebooks.
We argue that only through native support, testing in notebooks is going to be implemented in practice.

\textbf{Need for notebook-specific linting frameworks}. Perhaps, the most alarming result from our quantitative analysis is the rate of notebooks failing linting checks. Notebooks edited by experts are affected by an extensive number of issues, ranging from PEP~8 violations to actual warnings and errors. This confirms that code quality is a serious problem \cBlue{even when computational notebooks are written as a result of a collaborative effort and, as such,  are supposed to be clean, reproducible, and of high quality}~\cite{wang_better_2020}.
Code style violations are a serious concern, even for experienced notebook authors. As further evidence, during the interviews, BP9 (\textit{Stick to coding standards}) is mentioned in passing only by a couple of subjects. Yet, code style cannot be an afterthought, especially if the notebook itself, or the prototypical code developed therein, is targeted to production.

\textit{Implication for computational notebook platform designers}. We argue that the problem of poor code quality in notebooks might be due to the lack of a mature built-in linting tool for Jupyter notebooks. Fast and unobtrusive code checkers, providing readable suggestions on how to correct and refactor code -- like those available in modern IDEs -- would be greatly beneficial to improve the quality of notebooks while also taking into account their syntactic peculiarities.

\textbf{Limitations}.
\cTeal{
Although being specifically referred to Jupyter Notebook was not a selection criterion either for the articles retrieved in the MLR or the best practices that we extracted, we acknowledge that some of the items in our catalog address issues specific to the affordances and interface of the Jupyter platform. On the one hand, this is explained by the prominence we gave to Jupyter Notebook in the MLR search query; however,  a more general query with only `computational notebooks' would have likely resulted in missing many relevant entries for this study. On the other hand, Jupyter is the most popular notebook interface to date as well as the format to which many of the alternatives are inspired; as such, most of the modern literature on computational notebooks is reasonably focused on the Jupyter ecosystem.
}

Our qualitative analysis is based on a limited number of interviews with practitioners sampled opportunistically.
Nevertheless, it is worth noting that none of the interviewees mentioned best practices that had not been already identified through the literature review, thus increasing the confidence that we may have reached theoretical saturation, given the current state of the art and practice.
\cBlue{Regarding the interviewees, we only collected information about their educational background and the number of years working as data scientists at their company (see Table~\ref{tab:interviewees}).
However, we acknowledge that a better characterization and understanding of the interviewees' experience level would have allowed us to investigate whether and how expertise affects the way practitioners perceive or even implement best practices in their work routine.}

Our quantitative analysis is based on a dataset of Jupyter notebooks taken from Kaggle. We leveraged the Kaggle platform as it offered us a straightforward way to select high-quality and collaboratively developed notebooks. On the one hand, we believe that our sample is representative of the larger notebook population --- as the typical Kaggle challenge involves the usual activities for which notebooks are employed (e.g., explorative data analysis and prototypical modeling). On the other hand, we understand that our dataset may not cover all the possible notebook use cases.
\cBlue{For instance, in recent years, some companies have started leveraging computational notebooks in production systems.
One notable example is Netflix, which has developed an entire ecosystem of utilities (i.e.,  \texttt{papermill},\footnote{\url{https://github.com/nteract/papermill}} \texttt{commuter},\footnote{\url{https://github.com/nteract/commuter}} and \texttt{nteract})\footnote{\url{https://github.com/nteract/nteract}} to enable the scheduled execution of parametrized notebooks that are archived as historical records of analyses~\cite{ufford_beyond_2018, seal_part_2018}.
Beyond data analytics, computational notebooks can be used as software development tools for the creation of arbitrarily complex libraries, just like traditional IDEs. 
With \texttt{nbdev},\footnote{\url{https://nbdev.fast.ai/}} data scientists have access to a development environment based on Jupyter Notebook, which enables the creation of complete Python packages, including tests and documentation~\cite{howard_nbdev_2019}. 
By establishing a two-way synchronization between notebooks and source code, \texttt{nbdev} allows developers to switch back and forth from a traditional IDE  (e.g., for refactoring code) to Jupyter Notebook (e.g., to inspect the results of an execution).

Also related to the choice of mining data from Kaggle for our quantitative analysis, we acknowledge that the competitive nature of the platform may influence the adoption of some of the best practices from our catalog.
For example, \fullcite{cheng_building_2020} found that Kaggle expert users tend to share only partial, high-level solutions during the competitions, to preserve strategic details, and generally lack the time and will to expand their work once competitions are over, for the sake of knowledge sharing. Nonetheless, in our analysis, we did not take into account the notions of notebook completeness and level of abstraction, given that the focus of our study was specifically on verifying the adoption of the best practices in our catalog.
} In future work, we intend to replicate our study on different datasets, thus addressing any possible gap.

 \cBlue{
We also acknowledge that the motivations and personal goals are arguably different between the industry practitioners like our interviewees (e.g., career advancement, contributing to project success) and Kaggle experts (e.g., winning competitions, knowledge sharing). 
Nevertheless, we believe that the commitment of experienced data scientists to produce high-quality notebooks, motivated by the need to share them with peers and enable collaboration, provides a sufficient common ground for a complementary analysis of the two samples. 
}

\cBlue{Furthermore,} due to the peculiarities of the Kaggle platform, we were not able to define an operationalization for all the best practices in our catalog. 
For instance, we could not operationalize the best practices related to using version control and managing project-specific environments and dependencies (i.e., BP1, BP2, and BP3), as Kaggle enforces each of them by providing its users with a versioned and containerized notebook environment. 
\cBlue{We acknowledge this limitation, which we will address in future work by studying computational notebook platforms other than Kaggle.}

\cBlue{Concerning the operationalization of BP7 (\textit{Test your code}), we acknowledge that it might fail to detect testing libraries that are less known or domain-specific --- particularly when their names do not comprise the word `test' as a sub-string. Despite this possibility, we argue that the current operationalization was indeed useful for capturing the broad trend of the scarce use of testing in notebooks from Kaggle, thus confirming what \fullcite{pimentel_large-scale_2019} had already observed in their large-scale study on GitHub notebooks.}
Another limitation is related to the operationalization of expert data scientists as Kaggle users who have reached at least the expert tier in the platform. Although one may argue that the experience tiers in the platform may not reflect the actual expertise levels as data scientists, nowadays it is not uncommon for recruiters to use Kaggle and other platforms like Stack Overflow and GitHub \cite{marlow2013activity, sarma2016hiring} to identify job candidates using the status gained in the platform as a proxy for their actual expertise. Competitions are also often organized in Kaggle by employers looking for recruits among practitioners with strong skills in data science.\footnote{www.kaggle.com/c/about/host/recruiting} 

\cBlue{Moreover, we note that we complemented the quantitative analysis of the more `subjective' best practices --- whose operationalizations are potentially sub-optimal --- by performing a manual analysis of a large subsample of the dataset.}

In addition, the design of the Kaggle platform might have biased a couple of our findings. On the one hand, BP5 --- which recommends re-running notebooks top to bottom --- is the default behavior in Kaggle when submitting a notebook to a competition. On the other hand, the practice of importing local modules (part of BP6) relies on a feature that still lacks polish and is, therefore, rarely used.
We intend to address these limitations in future work by repeating this study on notebooks taken from different platforms (e.g., GitHub).

Finally, although we computed the related descriptive statistics, we could not operationalize opinionated best practices such as BP14 (\textit{Keep your notebook concise}): indeed, it is impossible to think of an operational definition of conciseness unless introducing the use of `magic numbers' lacking empirical validation.

Notwithstanding these limitations, we were able to assess the extent of adoption of most of the best practices from our catalog, especially the most controversial ones.

\section{Conclusion}
\label{sec:conclusion}

In this paper, we first performed a multivocal literature review to build  a catalog of 17 best practices for collaboration with computational notebooks; then, we qualitatively and quantitatively assessed the extent to which expert data scientists are familiar and comply with them. 

Overall, we found that data scientists are aware of the identified best practices and tend to adopt them in their daily work routine. Nevertheless, some of them have conflicting opinions on specific recommendations, due to varying contextual factors. Moreover, we found that the most neglected best practices have to do with code modularization, code testing, and adherence to coding standards. This highlights what we believe is a tension between speed and quality. 

To address this tension, computational notebook platforms should implement native features that ease the adoption of the aforementioned practices for collaboration. By having built-in, notebook-specific testing and linting frameworks as well as features for the refactoring and modularization of notebook code, we can expect data scientists to write high-quality notebooks without compromising on timeliness.


\begin{acks}
We would like to thank the anonymous interviewees.
We are also grateful to Michele Filannino, Fabio Fumarola, and Alexander Serebrenik for their help with the recruitment of volunteers.
\end{acks}

\bibliographystyle{ACM-Reference-Format}
\bibliography{references}

\received{January 2021}
\received[revised]{July 2021}
\received[accepted]{November 2021}

\newpage
\appendix

{\cBlueSimple
\section{Appendix A: Multivocal Literature Review}
}
\label{appendix:mlr}

\cBlue{To collect a catalog of best practices for the collaborative use of computational notebooks, we performed a Multivocal Literature Review (MLR). 
As a first step, we identified a set of suitable search terms. 
Next, we collected the search results and filtered them according to our research objective; then, we applied the snowballing technique to the selected articles. 
Lastly, we analyzed our final corpus, extracting best practices for collaboration around computational notebooks.
In this appendix, we describe the full process in detail.}

\cBlue{
\subsection{Search terms identification}
}

\cBlue{As a first step, we selected a set of suitable search terms for our MLR.
We started by building an initial set of terms composed of words that we had already encountered in the literature as well as their synonyms.
Then, we sought further search terms among the author-defined keywords from the results of a preliminary search. Lastly, we composed the final query.

\subsubsection{Initial set of search terms}

Our initial set of search terms was composed of the words: `computational notebook,' `Jupyter Notebook,' `best practice,' and `recommendation.' In the following paragraphs, we describe how we chose these terms and where we retrieved their synonyms.

\paragraph{`Jupyter Notebook'} As described in Section~\ref{sec:computational_notebook_platforms}, there are many possible implementations of a computational notebook.
We wanted our search results to be general enough and not to individually address a specific platform (e.g., Apache Zeppelin or  Wolfram Notebooks).
Nonetheless, we decided to explicitly include `Jupyter Notebook' as a search term since it is considered the most popular notebook platform among data scientists \cite{perkel2018jupyter} and, as such, it has attracted by itself most of the research on computational notebooks so far.
Moreover, the JSON-based file format defined by Jupyter Notebook (\texttt{.ipynb}) is the foundation for other similar file formats and is even directly supported by some of the Jupyter alternatives (e.g., Google Colab); furthermore, our focus is on identifying general-purpose recommendations that, if applicable to Jupyter Notebook, would also apply to other notebook development platforms.

\paragraph{`Best practice,', `Recommendation'} In \cite{pimentel_large-scale_2019}, the term `best practice' is used to present a list of final recommendations on notebook writing; we also included the term `recommendation' itself, as this is the expression used in the same paper to refer to the proposed best practices.

Additionally, before querying the digital libraries, we augmented the initial set of keywords by appending the synonyms of `recommendation.'\footnote{The term definitions and synonyms were retrieved from \cite{walter2008cambridge}.} 
Accordingly, we included the nouns `advice,' `tip,' `hint,' `guidance,' `guideline,' and `hack.'

\subsubsection{Refinement of the search terms}

To be sure to include all relevant search terms in our query, by following the example of Dillahunt et al.~\cite{dillahunt_sharing_2017}, we planned to iteratively search for relevant author-defined keywords in academic articles.

With the selected search terms, we performed a preliminary search in two full-text white literature databases, i.e., IEEE Xplore and ACM Digital Library.
Next, we reviewed each search result and identified relevant hits by inspecting paper titles and, in case of ambiguity, their abstracts.
Then, we extracted author-defined keywords from the subset of relevant hits and sought repeated keywords (i.e., occurring at least twice).
After filtering overly vague keywords (e.g., `exploratory programming') or conceptually unrelated (e.g., `automl'), we could not identify additional terms for our query.

\subsubsection{Definition of the final query}

Since we were not able to add any new term after the first search round, we stopped the iterative keyword search process there and built the final query for our MLR:}



\vspace{1mm}
{\cBlueSimple
\begin{lstlisting}
("best practice" OR "guideline" OR "recommendation" OR "advice" OR "tip" 
    OR "hint" OR "guidance" OR "hack") 
AND 
("Jupyter Notebook" OR "Jupyter notebooks" OR "computational notebook")
\end{lstlisting}
}
\vspace{1mm}

\begin{figure}
    \centering
    \includegraphics[width=\textwidth]{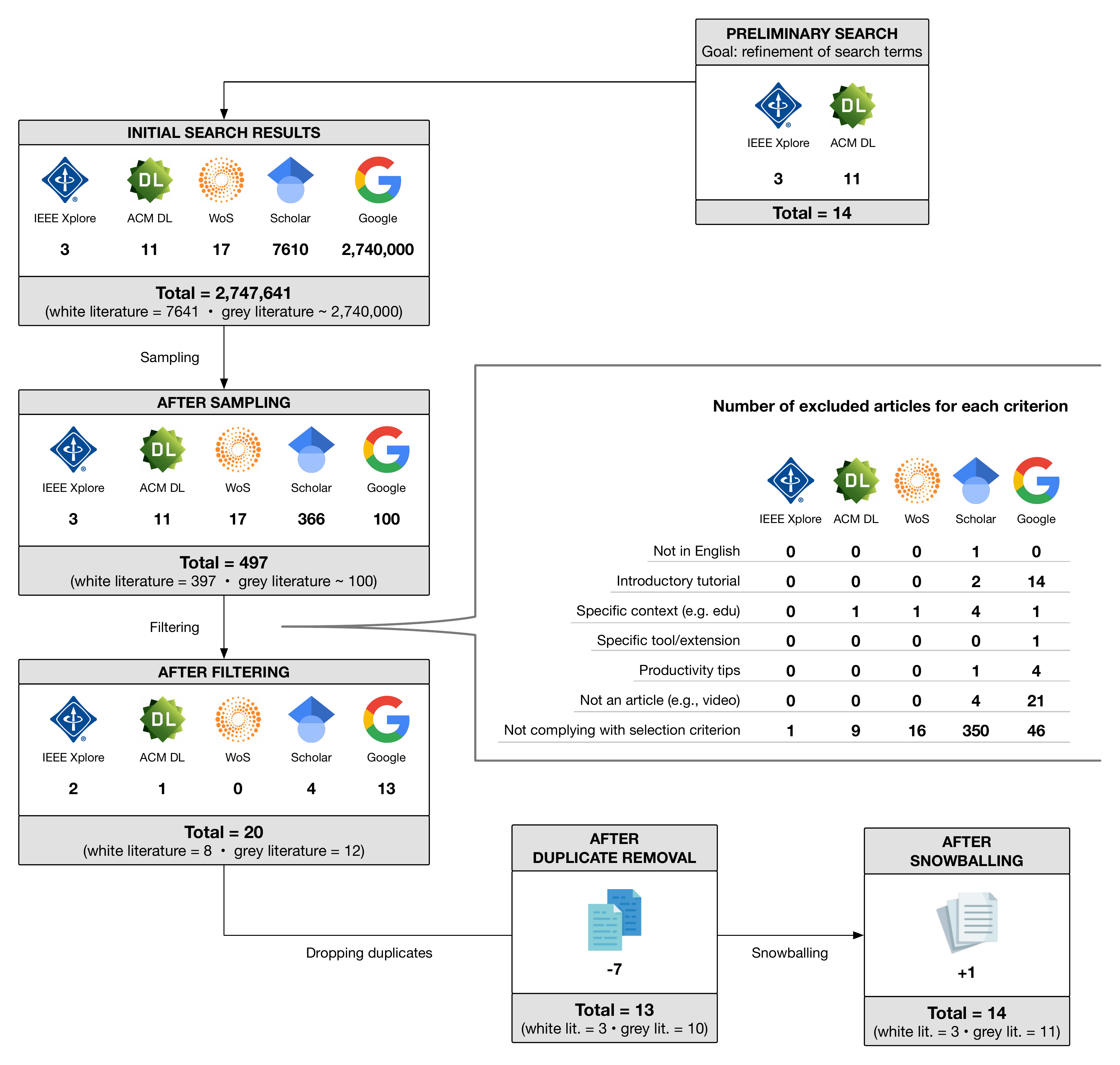}
    \cBlueCaption
    \caption{Infographic summarizing the data collection process.}
    \label{fig:search-results-infographic}
\end{figure}

\cBlue{
Note that the query shown here is a simplified version of the one used in our study; in the actual query, we explicitly included countable search terms in both their singular and plural form.
Moreover, the simplified query reported here is composed using a generic syntax; in its final form, the query has been specialized to account for the syntactic requirements of each specific search engine.

\subsection{\cBlue{Data Collection}}

To retrieve white literature (i.e., peer-reviewed scientific papers), we queried four digital libraries: in particular two broad-coverage abstract databases, i.e., Google Scholar and Web of Science, and two limited-coverage full-text databases, i.e., IEEE Xplore and ACM Digital Library.
These libraries comprise a broad range of scientific publication types, including books or papers published in journals, conference proceedings, and technical magazines.
Instead, to retrieve grey literature, we performed a generic web search using Google.

By querying the selected search engines, we retrieved a total of 7,641 white literature hits and 2,740,000 grey literature hits.
Details on the number of search results retrieved from each search engine are reported in Figure~\ref{fig:search-results-infographic}.

\subsubsection{Sampling criteria}

All academic digital libraries returned a small number of search hits, which we completely and thoroughly reviewed (IEEE Xplore, ACM Digital Library, and Web of Science returned 3, 11, and 17 results respectively); the only exception was Google Scholar, which returned \numprint{7620} results.
In this case, we went through the first 366 search hits (i.e., 37 10-item pages) --- the same size of a representative sample of the complete result set with a 5\% margin of error and 95\% confidence level.

The generic search on Google returned an extensive number of results (2,740,000).
We checked the relevance of the first 100 hits (i.e., ten 10-item pages), and stopped at the 10th page after inspecting six consecutive pages of unrelated results.

\subsubsection{Selection criteria}

The selection and exclusion criteria that we used to decide whether to accept each search result are summarized in Table~\ref{tab:search-results-criteria}, along with their rationale.

\begin{table}[t]
\cBlueSimple
\cBlueCaption
\small
\caption{List of inclusion and exclusion criteria applied to filter search results in the MLR}
\label{tab:search-results-criteria}
\begin{tabular}{p{7cm}p{6.5cm}}
\toprule

\multicolumn{2}{c}{\textbf{Search results inclusion criterion}} \\ 

\multicolumn{2}{p{13.5cm}}{The search result contains best practices for the collaborative use of computational notebooks in a generic professional context.} \\

\toprule
\multicolumn{2}{c}{\textbf{Search results exclusion criteria}} \\

\multicolumn{1}{c}{\textbf{Criterion}} & \multicolumn{1}{c}{\textbf{Rationale}} \\ \toprule

    The article was published before 2015.
    &
    Project Jupyter was officially formed in Feb 2015.\\ \midrule

    The article is not written in the English language.
    &
    Constraints of our research team; high-quality research in computing is generally published in English.\\ \midrule

    The article is an introductory tutorial on computational notebooks.
    &
    Our focus is on best practices for the collaborative use of notebooks in a generic professional context.\\ \midrule

    The article presents best practices for the use of notebooks in a specific context/domain (e.g., the use of Jupyter notebooks in education, as in \cite{johnson_benefits_2020}).
    &
    Same as above.\\ \midrule

    The article presents or advertises a specific tool or notebook extension (e.g., \cite{wang_restoring_2020}).
    &
    Our focus is on best practices in terms of optimal behaviors (rather than optimal tooling), to be adopted while collaboratively using notebooks.\\ \midrule

    The article merely provides productivity tips for writing notebooks.
    &
    Our focus is on optimal behaviors for better collaboration with notebooks, not on tips  \& tricks for faster notebook writing.\\ \midrule

    The search result is not an article or a blog post (e.g., it is a video, podcast, or discussion thread, etc.). 
    & -
    \\ \bottomrule
    
\end{tabular}
\end{table}

We included search results containing best practices for the collaborative use of computational notebooks in a generic professional context.
We decided to keep only search results consisting of articles or blog posts written in English, thus excluding other kinds of multimedia (e.g., video, podcasts, discussion threads, etc.).
The language limitation is partly due to the educational constraints of our research team and partly motivated by the consideration that high-quality research in computing is generally published in English \cite{dillahunt_sharing_2017}.
Moreover, we wanted to collect items concerning recommendations for the use of modern notebook development platforms: accordingly, we filtered out articles published before 2015, as they would have predated the launch of Project Jupyter and, as a consequence, the release of many other Jupyter-inspired modern notebook applications (e.g., Google Colab). This filtering criterion was applied directly when issuing the final query in the search engines, by adding the condition $\geq$~\texttt{2015/01/01} on the publication date fields.

Being interested in best practices for the collaborative use of notebooks in a generic professional context, we excluded introductory tutorials on computational notebooks. 
For the same reason, we kept out articles on the use of notebooks in a specific context or domain, such as those pertaining to the use of notebooks in education.
Indeed, by inspecting a sample of them, we realized that they typically contain recommendations that are hard to generalize to a professional context.

We also excluded search results merely providing low-level tips for better productivity in notebook writing or opinionated articles recommending the use of a specific tool or notebook extension.
These articles generally miss the main focus of our research, that is, the identification of optimal (high-level) behaviors for better collaboration with computational notebooks.

The application of the filtering criteria described so far to the search results retrieved from each search engine left us with a final corpus of 16 papers (not considering duplicates).

\subsubsection{Snowballing}

Given the low number of selected papers, we employed the snowballing technique to find further academic articles that we might have missed during the automated search.
In particular, we engaged in the backward and forward snowballing process \cite{wohlin_guidelines_2014} by leveraging respectively  the references within and the citations of the selected white literature.

Our start set included the only three academic papers that we had selected \cite{rule_ten_2018, pimentel_large-scale_2019, beg_using_2021-1}. We did not retrieve any new article with the forward snowballing strategy, despite the high number of citations that the first two seed papers had received, i.e., 45 and 42 respectively; conversely, \cite{beg_using_2021-1} was cited only two times. With backward snowballing, we found no further references from \cite{pimentel_large-scale_2019} and \cite{beg_using_2021-1}, and only one relevant reference \cite{woodbridge_jupyter_2017} from \cite{rule_ten_2018}. This new reference is a grey literature article (i.e., a blog post) containing only two white literature references, one conceptually unrelated to our research objective and the other already included in our corpus \cite{rule_ten_2018}. Therefore, we interrupted the process after the first snowballing iteration.

This was the last step of our data collection process, which is visually summarized in Figure~\ref{fig:search-results-infographic}.

}

\subsection{\cBlue{Data Extraction}}

\cBlue{We carefully read each of the selected articles; in doing so, we extracted excerpts of the given recommendations and gathered a raw catalog of best practices. 

By `best practice' we specifically mean an optimal, high-level, and general behavior that can be adopted to improve collaboration around computational notebooks. Such behavior should be general enough to be implemented in modern computational notebook platforms in general, i.e., it should be neither platform-specific nor language-specific.
For these reasons, while transcribing recommendations, we used the inclusion and exclusion criteria reported in Table~\ref{tab:best-practices-criteria} to filter out those recommendations lacking generality, objectivity, or only suggesting the use of third-party, platform-specific extensions.

Furthermore, we double-checked that each recommendation was related to behaviors with a direct or indirect impact on collaboration. Thus, we discarded best practices that were clearly addressed to the workflow of a single analyst (e.g., ``store/cache a serialized version of your intermediate results/outputs'').}

{\cBlueSimple
\subsection{Data Analysis}
}

\cBlue{Finally, to further refine our catalog, we performed a thematic analysis of the raw collection of best practices \cite{clarke2015thematic}.
In particular, we assigned a set of open codes to each best practice and, upon consolidation of the code set, we performed a final step of focused coding.
With each code, we tried to summarize a set of excerpts containing similar recommendations; therefore, the codes themselves can be read as high-level recommendations.
Moreover, for each identified code, we noted down its support (i.e., the number of sources in which we observed it).

The open coding session was carried out by the first author of the paper; the resulting set of codes was discussed and refined with the help of the second author. The focused coding session was then performed in parallel by the two authors; in case of disagreements, the final codes were determined by all the three authors of the paper.
Finally, to facilitate the discussion, we grouped codes into themes.}

\begin{table}[tbh]
\cBlueSimple
\cBlueCaption
\small
\caption{List of inclusion and exclusion criteria applied to filter best practices in the MLR}
\label{tab:best-practices-criteria}
\begin{tabular}{p{7cm}p{6.5cm}}

\toprule

\multicolumn{2}{c}{\textbf{Best practices inclusion criterion}} \\ 

\multicolumn{2}{c}{The article excerpt describes a best practice for the collaborative use of computational notebooks.} \\

\toprule

\multicolumn{2}{c}{\textbf{Best practices exclusion criteria}} \\ 
\multicolumn{1}{c}{\textbf{Criterion}} & \multicolumn{1}{c}{\textbf{Rationale}} \\
\toprule

    The best practice merely recommends the use of one or more Jupyter Notebook extensions (e.g., \texttt{toc}).
    &
    Our focus is on best practices in terms of optimal behaviors (rather than optimal tooling), to be adopted while collaboratively using notebooks.\\ \midrule

    The best practice recommends the use of external tools (e.g., full-fledged IDEs like PyCharm).
    &
    Same as before. \\ \midrule

    The best practice recommends the use of specific IPython magic functions (e.g., \texttt{\%run}), or python packages (e.g., \texttt{scikit-learn}), rather than a more generalizable behavior.
    &
    Our focus is on high-level recommended behaviors; such kinds of recommendations have too low a level of abstraction. \\ \midrule

    The best practice is merely a productivity tip (e.g., \textit{``learn the Jupyter Keyboard Shortcuts''}, or \textit{``create a notebook template with the import statements that you use the most''}, etc.).
    &
    Our focus is on optimal behaviors for better collaboration with notebooks, not on tips \& tricks for faster notebook writing.\\ \midrule

    The best practice is an opinionated recommendation not supported by any argumentation (e.g., \textit{``use Jupyter Lab instead of Jupyter Notebook''}, \textit{``use Python 3 instead of Python 2''}, or \textit{``develop locally and execute remotely''}).
    &
    Best practices are necessarily subjective and none of them can count on unanimous acceptance among professionals; nevertheless, we decide to exclude opinionated recommendations when they lack a proper argumentation. \\ \midrule

    The best practice is a generally valid coding best practice (e.g., \textit{``Use meaningful variable names''}, or \textit{``Keep the code DRY''}) and is not notebook-specific; the best practice is a recommendation for code optimization (e.g., \textit{``Avoid loops and prefer array operations''}).
    &
    Again, these kinds of recommendations have too low a level of abstraction; moreover, our focus is on notebook writing, not on code writing. \\ \bottomrule
    
\end{tabular}
\end{table}

{\cBlueSimple
\section{Appendix B: Manual coding}
\label{appendix:coding}
}

\cBlue{To perform our quantitative analysis of Kaggle notebooks, we found that some of  the best practices in our catalog could only be operationalized partially.
Specifically, to understand if a notebook is well documented (BP11 \textit{Document your analysis}) or if the headings in its MD cells actually contribute to give it a proper structure (BP12 \textit{Leverage Markdown headings to structure your notebook}), the established quantitative measures provided us with a partial picture. 
In addition, concepts like notebook cleanliness (BP13) and conciseness (BP14) are difficult to evaluate by solely relying on a set of numeric measures. 
For this reason,  we engaged in the manual coding of a representative sample of Kaggle notebooks to complement the results from the quantitative analysis of the  four best practices BP11-14 and further assess the extent of their adoption.}

{\cBlueSimple
\subsection{Method}
}

\cBlue{First, we randomly selected a representative sample of 301 items from the complete dataset of 1,380 notebooks used for the quantitative analysis.
Next, the first two authors of the paper independently engaged in a pilot open-coding session; after analyzing 10 notebooks, the authors met and discussed the set of open codes they had come up with, reconciling any discrepancies. 
Once the full agreement was reached, for each best practice we set the codes in the form of a  3-point Likert scale, namely \textit{Poor}, \textit{Fair}, and \textit{Good}, to gauge the extent to which the notebooks would comply with it.
Also, to facilitate the coding, the authors wrote an operational definition for each code; the list of such operational definitions completed the coding schema used to annotate the rest of the notebooks (see Table~\ref{tab:manual-coding-codes-defs}).}

\begin{table}[tb]
\cBlueSimple
\cBlueCaption
\small
\caption{Coding schema.}
\label{tab:manual-coding-codes-defs}
\begin{tabular}{@{}lp{12cm}@{}}
\toprule
\multicolumn{2}{c}{\textbf{Compliance to BP11 (\textit{Document your analysis})}}                                               \\ \midrule
Poor &
    MD cells are not present in the notebook, or they are not used to convey decision rationales and to explain the output of the analyses. \\
Fair &
    MD cells are generally used to convey decision rationales and to explain the output of the analyses, although explanations are kept to a decent minimum. \\
Good &
    MD cells provide a thorough explanation of decision rationales and present a detailed discussion of the output of the analyses. \\ \midrule
  
\multicolumn{2}{c}{\textbf{Compliance to BP12 (\textit{Leverage MD headings to structure your notebook})}}                           \\ \midrule
Poor &
    MD headings serve other purposes than structuring the notebook or they are not used at all.      \\
Fair &
    MD headings give some structure to the notebook, although they are not always used consistently.    \\
Good &
    MD headings are used to structure the notebook consistently throughout its length.                  \\ \midrule

\multicolumn{2}{c}{\textbf{Compliance to BP13 (\textit{Keep your notebook clean})}}                                             \\ \midrule
Poor &
    The notebook contains many empty cells, unexecuted cells, or corrupted output cells (e.g., error stack traces, empty or corrupted images, etc.). \\
Fair &
    Overall, the notebook looks clean, although we observe some minor defects.                         \\
Good &
    The notebook looks perfect (here we only forgive one empty cell at the end).                        \\ \midrule

\multicolumn{2}{c}{\textbf{Compliance to BP14 (\textit{Keep your notebook concise})}}                                           \\ \midrule
Poor &
    The notebook or some of its cells are too long and have multiple responsibilities.                  \\
Fair &
    Overall, the notebook and its cells are concise, although we observe one or more cells that might be better shortened, as they are too long and have multiple responsibilities. \\
Good &
    The notebook and its cells are self-contained and have a single responsibility, without exceptions. \\ \bottomrule
\end{tabular}
\end{table}

\cBlue{After the definition of the coding schema, the first two authors coded independently the first 90 notebooks from the random sample and calculated the Cohen's $\kappa$ coefficient to evaluate the inter-rater reliability for each of the four best practices under assessment.
The agreement level ranged from substantial, for  BP13 ($\kappa = 0.79$) and BP14 ($kappa = 0.78$), to  strong, for BP11 ($\kappa = 0.86$ and BP12 ($\kappa = 0.87$)~\cite{landis1977measurement}.
Then, the authors met to reconcile the disagreements. 
We point out that no further refinement of the coding schema was necessary and that no cases of strong disagreements were observed (i.e., rater A chose \textit{Poor} while rater B chose \textit{Good} or vice versa).
The rest of the sample was then coded by the first author of the paper.}

{\cBlueSimple
\subsection{Results}
}

\begin{table}[tb]
\cBlueSimple
\cBlueCaption
\small
\caption{Results from the manual notebooks coding.}
\label{tab:manual-coding}
\begin{tabular}{@{}lrrr@{}}
\toprule
 \textbf{Best practice}    & \multicolumn{1}{l}{\textbf{\textit{Poor
}}} & \multicolumn{1}{l}{\textbf{\textit{Fair}}} & \multicolumn{1}{l}{\textbf{\textit{Good}}} \\ \midrule
BP11 \textit{Document your analysis} & 64.63\%                  & 20.73\%                  & 14.63\%                  \\
BP12 \textit{Leverage Markdown headings} & 39.02\%                  & 7.32\%                   & 53.66\%                  \\
BP13 \textit{Keep your notebook clean} & 4.88\%                   & 14.63\%                  & 80.49\%                  \\
BP14 \textit{Keep your notebook concise} & 18.29\%                  & 21.95\%                  & 59.76\%                  \\ \bottomrule
\end{tabular}
\end{table}

\cBlue{In Table~\ref{tab:manual-coding} we report the code frequencies for the guidelines under assessment.
Regarding BP11 (\textit{Document your analysis}), although the majority of the notebooks contain MD cells (83.12\%), in most cases the documentation was coded as \textit{Poor} (64.63\%) and judged \textit{Good} only in the 14.63\% of the notebooks. As for BP12 (\textit{Leverage Markdown headings}), in 55.66\% of the cases, MD headings were found to properly structure the notebooks, while in 39.02\%  the use of headings was judged as \textit{Poor}.
As far as the notebook cleanliness is concerned (BP13), most of the notebooks (80.49\%) were coded as \textit{Good}.
Finally, notebook conciseness (BP14) was found \textit{Good} in 59.76\% of the notebooks, \textit{Fair} in 21.95\%, and \textit{Poor} in the remaining 18.29\%.}

\end{document}